\documentclass[11pt,journal,letterpaper,compsoc,onecolumn,draftclsnofoot]{IEEEtran}
\pdfoutput=1

\input{macros}

\begin{document}

\title{Model Checking Software Programs with First Order Logic Specifications using
AIG Solvers}

\author{Fadi A. Zaraket, Mohamad Noureddine
\IEEEcompsocitemizethanks{
\IEEEcompsocthanksitem Fadi A. Zaraket is an Assistant Profesor with the
Department
of Electrical and Computer Engineering, American University of Beirut,
Beirut, Lebanon.\protect\\
E-mail: {\texttt fadi.zaraket@aub.edu.lb}
\IEEEcompsocthanksitem M. Noureddine is a graduate student at the
Department
of Electrical and Computer Engineering, American University of Beirut, Beirut,
Lebanon.\protect\\
E-mail: {\texttt man17@aub.edu.lb}
}% <-this % stops a space
\thanks{}}

\IEEEcompsoctitleabstractindextext{
\begin{abstract}
Static verification techniques leverage Boolean formula satisfiability solvers
such as SAT and SMT solvers that operate on conjunctive normal form and first 
order logic formulae, respectively, to validate programs. 
They force bounds on variable ranges and execution time and 
translate the program and its specifications into a %conjunctive normal form 
Boolean formula. 
They are limited to programs of relatively low complexity for the following reasons. 
(1) A small increase in the bounds can cause a large increase 
in the size of the translated formula.  
(2) Boolean satisfiability solvers are restricted to using optimizations that apply 
at the level of the formula.
Finally, (3) the Boolean formulae often need to be regenerated with higher bounds 
to ensure the correctness of the translation.

We present a method that uses sequential circuits, Boolean formulae with memory 
elements and hierarchical structure, and sequential circuit synthesis 
and verification frameworks to validate programs. 
(1) Sequential circuits are much more succinct than Boolean formulae with no memory
elements and preserve the high-level structure of the program.
(2) Encoding the problem as a sequential circuit enables the use of a number 
of powerful automated analysis techniques that have no counterparts for 
other Boolean formulae. 
Our method takes an imperative program with a first order logic specification 
consisting of a precondition and a postcondition pair, and a bound on the
program variable ranges, and produces a sequential circuit with a designated
output that is \true when the program violates the specification. 
Our method uses sequential circuit synthesis reduction techniques to reduce the 
generated circuit, and then uses sequential circuit verification techniques to 
check the satisfiability of the designated output. 
The results show that our method can validate designs that are not 
possible with other state of the art techniques, and with bounds that are an 
order of magnitude larger.% than those achievable by the state of the
%art techniques.
%based on conjunctive normal form analysis.
\end{abstract}

}

\maketitle

%\listoftodos

%\clearpage
%%%%%%%%%%%%%%%%%%%%%%%%%%%%%%%%%%%%%%%%%%%%%%%%%%%%%%
%%%%%%%%%%%%%%%%%%%%%%%%%%%%%%%%%%%%%%%%%%%%%%%%%%%%%%

\section{Introduction}
\label{s:intro}
%% Introduction
The work in~\cite{SERA07} takes a declarative formula $\phi$ in first order logic
(FOL) with transitive closure and a bound on the universe of discourse and
translates it to a %n equisatisfiable 
sequential circuit expressed in VHDL. 
A sequential circuit is a Boolean netlist with memory elements and a hierarchical
structure. 
It then passes the sequential circuit to 
SixthSense~\cite{Hari05expert}, an IBM internal sequential circuit verification 
framework,
and decides the validity of $\phi$ within the bound. 
It scales to bounds larger than what is possible with 
Kodkod~\cite{kodkodTJ2007} which translates $\phi$ into a propositional Boolean 
formula in conjunctive normal form (CNF) and checks its validity using a 
Boolean satisfiability solver such as MiniSat~\cite{sorensson2005minisat}.

The work in~\cite{SEBAC07} translates an imperative C program, with an assertion
statement therein, and a bound on the input size, into a sequential circuit expressed 
in VHDL. 
It then passes the sequential circuit to SixthSense~\cite{Hari05expert}
and decides the validity of the assertion within the bound. 
It scales to bounds larger than what is possible with CBMC~\cite{clarke2004tool}.
CBMC unwinds loops and recursive functions up to the given bound, 
and translates the program 
into a CNF formula that asserts the properties. 
It then checks for correctness using a Boolean satisfiability solver. 

In this work, we present our method that takes an imperative program \Pm
with a specification, FOL precondition and postcondition pair \pair{\Pre}{\Post},
and checks whether \Pm satisfies the specification within a bound $b$ on the 
domain of the program and specification variables (\prob);
i.e. when the bounded inputs of \Pm satisfy \Pre, the outputs of \Pm satisfy \Post. 
%The postcondition relates the program's outputs to its inputs~\cite{bradley2007calculus}. 
The program is written in \psqlanguage, a subset of C++/Java that includes 
integers, arrays, loops, and recursion.
%The specification is written in FOL. 
%consists of a precondition \Pre and a postcondition \Post and 
%both are written in FOL. 
Our method translates the problem \prob into a sequential circuit with a designated 
output therein that is \true iff the program violates the specification
within the bounded domain.
Our method uses sequential circuit synthesis reduction and abstraction techniques 
embedded in the open source ABC~\cite{brayton2010abc} framework 
to reduce the generated sequential circuit. 
Then it uses ABC verification techniques to decide the satisfiability of the designated
output.
ABC either (1) proves the validity of the program, 
(2) generates a counterexample illustrating that the program violates the 
specifications, 
or (3) reports an inconclusive result as it exhausts computational resources. 
Our method translates the counterexample back to 
the program domain and provides the user with a visual debugging tool 
(GTKWave~\cite{gtkwave}) to trace the violation. 

Our method significantly extends the work of both \cite{SERA07,SEBAC07} in that
\be

\i it uses a program counter semantics to translate the program into an intermediary
{\em one loop program} (\oneloop), 
where a program counter is an additional variable that encodes the control flow of 
the program. 
It encodes the data flow of the program into ternary conditional statements based on 
the value of the program counter.  
The work in \cite{SEBAC07} performs only a source to source translation to VHDL, 

\i it directly translates the \oneloop program into bit level representation using {\em
And-Inverter-Graphs} (AIG) sequential circuits while \cite{SERA07,SEBAC07} depend on 
the VHDL compiler and synthesis tools to do the translation to bit level, 

\i it supports imperative programs annotated with FOL specifications and also 
supports array boundary and overflow checks, 

\i it supports function calls including recursion, %and requires a bound on recursion depth only if the recursive function uses local variables, 

\i in case the original correctness check was not conclusive, it uses heuristics 
to guess a termination bound on the program execution time, 
it enables a termination guarantee check within the execution time bound, and 
it then uses the execution time bound with bounded model checking to
decide correctness,

%\i it provides a simulator that generates nondeterministic values for inputs and 
%simulate the program, 

\i it uses ABC~\cite{brayton2010abc}, an open source sequential circuit synthesis 
and verification framework, instead of SixthSense~\cite{Hari05expert} an IBM 
internal sequential circuit solver.
ABC is a transformation-based verification (TBV)~\cite{brayton2010abc} framework that 
operates on sequential circuits and iteratively and synergistically 
calls numerous reduction and abstraction algorithms such as 
retiming~\cite{KuBa01}, redundancy 
removal~\cite{HmBPK05,KuMP01,BjesseC00,aziz-fmsd-00}, logic
rewriting~\cite{BjBo04}, interpolation~\cite{McMillan03}, 
and localization~\cite{Wang03}.
These algorithms simplify and decompose complex problems 
until they become tractable for ABC verification techniques such 
as symbolic model checking,
bounded model checking, induction, interpolation, 
circuit SAT solving, and  target
enlargement~\cite{MoGS00,MoMZ01,HoSH00,BaKuAb02,Hari05expert},

\i our method is fully implemented as an open source tool (\psqtool) available online
~\footnote{\label{fn:online}\url{http://research-fadi.aub.edu.lb/dkwk/doku.php?id=sa}}.

\ee

We evaluated our method with the verification of 
standard algorithms, fundamental and complex data
structures, real applications and programs from the software verification 
benchmarks and compared our method to CBMC and other tools ranking top in the 
software verification competition~\cite{SVComp2014}.
\psqtool succeeded to find and report counterexamples for all defected programs.
It found and reported defects that we were not aware of while developing the
evaluation benchmarks.
It scaled to verification bounds higher than those possible with the other tools,
and proved specifications that were not possible by the other tools.

\subsubsection*{Limitations of translation to Boolean formulae}
\label{s:intro:limitations}
Recent advances in SAT enabled tools like Kodkod Alloy Analyzer\cite{kodkodTJ2007},
and CBMC~\cite{clarke2004tool}
%EBMC, and VCC 
to check real programs.
However, these programs often need to be partial, leaving out important functionality 
aspects, to enable the analysis to complete.  
Moreover, the analysis is typically bound to relatively small limits. 
%e.g., fewer than 7 nodes in a tree structure with KodKod.

%\section{Comparison with recent work}
%Tools such as CBMC~\cite{clarke2004tool}
%check for pointer safety, within bound
%array access and user defined assertion properties in C programs.
%Given a C program and a bound on the range of variables, CBMC unwinds the program's 
%loops and recursive functions, and unfolds the program 
%into a Boolean (CNF) formula that asserts the specified properties. 
%It then uses SAT methods and tools such as MiniSat~\cite{sorensson2005minisat}
%%to check the CNF formula for counter examples. 
%%As the given bound increases, and the program becomes more complexe, the size
%%of the generated CNF formula rapidly grows and SAT problem becomes practically
%%impossible to solve. Moreover, programs containing unbounded loops might 
%%even render the process of generating the CNF formula infeasible. 

There are three limiting aspects of translating high-level programs to Boolean 
formulae.  
\begin{LaTeXdescription}
\item[Disadvantage 1]
The translation to Boolean formulae depends on the bounds; a small increase in 
the bound on variable ranges can cause a large increase in the size of the
translated formula due to unwinding loop and recursion
structures in programs, or eliminating quantifiers 
%and unrolling transitive closure 
in declarative first order logic.
%for
%example, for an undirected seven-node tree the translation
%from Alloy to CNF generates a formula with over one million
%variables and five million clauses.  

\item[Disadvantage 2] 
CNF SAT solvers are restricted to
using optimizations, such as symmetry
breaking~\cite{Aloul02SymSAT} and observability don't
cares (ODC)~\cite{FuYuMalik2005}, that apply at the level
of CNF formulae.  
However these optimizations usually aim
at increasing the speed of the solver and often result in
larger formulae as they add literals and clauses to the
CNF formula to encode symmetry and ODC
optimizations~\cite{ZhuKu06SATSweepODC}.  
Often times when the analyzer successfully generates a large CNF formula,
the underlying solver requires intractable resources.

\item[Disadvantage 3] 
Often times the formula needs to be regenerated with higher bounds in
case the unwinding bounds were not large enough for the
loops to complete as is the case with CBMC and ESBMC. 
%Note that 
%Multiple bounds exist and they need not be all
%increased during one iteration.
\end{LaTeXdescription}

To extend the applicability of static analysis to a wider class of programs as 
well as to check more sophisticated specifications and gain more confidence in the
results, we need to scale the analysis to significantly
larger bounds. %limits on the range of program variables.

\subsubsection*{Advantages of sequential circuits}
\label{s:intro:advantages}
We formally define sequential circuits in
Section~\ref{sec:sequential}; for now a sequential
circuit can be viewed as a restricted C++ program,
specifically a concurrent program in which all
variables are either integers, whose range is statically
bounded, or Boolean-valued, and dynamic allocation is
forbidden~\cite{edwards2005challenges}.
%
%Given a program with a specification and a bound, we
%automatically derive a sequential circuit and a designated Boolean output 
%variable therein that serves as an {\em invariant}, i.e.,
%the output is $\mbox{true}$ if and only if the
%program violates the specification within the bound. 
%
There are two key advantages to compiling programs 
into sequential circuits rather than Boolean formulae:
\begin{LaTeXdescription}
\item[Advantage 1] 
Sequential encodings are much more succinct than SMT or pure combinational 
SAT formulae.
They are imperative and state-holding while CNF and SMT formulae are 
declarative and state-free.  
For example, they can naturally represent the execution of quantifiers and 
loops without the need for unrolling them.
Moreover, they can store and reuse intermediate results in local variables.
In cases, SAT and SMT encoding algorithms produce a data structure that uses 
several orders of magnitude more memory to represent.

\item[Advantage 2]  
Casting the decision problem for a program specification as an invariant check 
on a sequential circuit allows to leverage a number of powerful 
automated analysis techniques that we discuss in Section~\ref{s:back:abc}
and that have no counterpart in CNF or SMT analysis. 
\end{LaTeXdescription}

Other software verification techniques and tools exist that leverage 
predicate abstraction, interpolation, model checking, SMT,  
and other FOL and CNF solvers~\cite{Arm06boundedmodel,SATABS2005,VCC2009TPHOLS,jpfVisser03visser,sem96,darwinfm2007,Xie03FME}. 
We discuss the tools and further compare our method 
to them in Section~\ref{s:related}. 

%In this work we make the following contributions. 
%\be
%\i We encode an imperative program \Pm with a first order logic 
%specification \pair{\Pre}{\Post} into a sequential circuit with an invariant
%that is stuck to $\mathit{true}$ iff  \Pm satisfies \Post for all inputs that
%satisfy \Pre within a given bound on the range of the program variables. 
%\i We use the ABC sequential circuit verification framework to check the
%circuit and we check formulae and programs that are orders of magnitude
%higher than those possible with the Alloy Analyzer and the CBMC tool. 
%\i We use a program counter semantics to encode the \Pm and
%\pair{\Pre}{\Post} into a sequential circuit. 
%We use that to encode and compute
%a termination guarantee check within a bound on the number of
%iterations or recursive calls a program can make. 
%We use the termination guarantee bound to prove run time efficiency of given
%algorithms.
%We also use the termination guarantee bound to prove conclusive correctness
%using bounded model checking, in case other proof techniques did not return
%conclusive results. 
%\i We provide our tool and benchmarks online$^{\ref{fn:online}}$. 
%\i visual debugger \todo{gtkwave... }
%\ee

%\itodo{make a statement that this dissertation is not a statement against SAT since simple TBV uses SAT}

The rest of this paper is structured as follows.
Section~\ref{s:back} reviews
basic definitions and concepts related to Boolean
formulae, and introduces the \psqcore and \psqlanguage programs, 
sequential circuits, and the ABC framework.
Section~\ref{s:overview} provides an overview of the \psqtool, introduces
\oneloop programs, and illustrates the method with an array search example. 
Section~\ref{s:transformation} describes the translation of \psqlanguage programs
into sequential circuits. 
%Section~\ref{s:termination} discusses the termination guarantee bound computation. 
Section~\ref{s:implementation} discusses the implementation.
We discuss the results in Section~\ref{s:results}, the related work in
Section~\ref{s:related}, and conclude with future work in
Section~\ref{s:conclusion}.

%%%%%%%%%%%%%%%%%%%%%%%%%%%%%%%%%%%%%%%%%%%%%%%%%%%%%%
%%%%%%%%%%%%%%%%%%%%%%%%%%%%%%%%%%%%%%%%%%%%%%%%%%%%%%

\section{Background}
\label{s:back}
In this section we define programs, sequential circuits, and 
introduce the ABC synthesis and verification framework. 
A reader well-versed in software verification may wish 
to skip this section, using it only as a reference.

\begin{lrbox}{1}
\begin{tabular}{p{11cm}p{10cm}}
\lstset{
   keywords=[1]{declaration-statement,declaration,program,list-of-statements,statement,conditional,loop,sync,assignment,target,boolean-expr,expression,return-statement,type,variable-decl,function-decl,property-decl,precondition,postcondition,property,quantifier,range,term,access,function-call,call-argument,argument-list,call-arg-list,var,array-var,array-decl,then-block,else-block,while-block,target-term,array-access},
   keywords=[2]{@dotogether,@pre,@post,if,while,return,forall,exists,int,bool,num,else,constant,a-un-op,a-bin-op,b-un-op,b-bin-op,ba-bin-op,id},
}
\begin{lstlisting}
program: declaration-statement+ statement+
statement: assignment | conditional | loop
list-of-statements: statement*

// statements
assignment: target /*@\BoldDarkRed{=}@*/ expression /*@\BoldDarkRed{;}@*/
conditional: if /*@\BoldDarkRed{(}@*/boolean-expr/*@\BoldDarkRed{)}@*/ /*@\BoldDarkRed{\{}@*/ then-block /*@\BoldDarkRed{\}}@*/ else /*@\BoldDarkRed{\{}@*/ else-block /*@\BoldDarkRed{\}}@*/
loop: while /*@\BoldDarkRed{(}@*/ boolean-expr /*@\BoldDarkRed{)}@*/ /*@\BoldDarkRed{\{}@*/ while-block /*@\BoldDarkRed{\}}@*/
then-block, else-block, while-block: list-of-statements

// declarations
declaration-statement: declaration /*@\BoldDarkRed{;}@*/
declaration: (variable-decl | array-decl)
variable-decl: type var 
array-decl: type array-var /*@\BoldDarkRed{[}@*/ constant?  /*@\BoldDarkRed{]}@*/ (/*@\BoldDarkRed{[}@*/ constant /*@\BoldDarkRed{]}@*/)?
var, array-var : id
type: int | bool            

// expressions
boolean-expr: term | b-un-op boolean-expr | boolean-expr b-bin-op boolean-expr | expression ba-bin-op expression
term: constant | target-term 
target-term: var | array-access
array-access: array-var /*@\BoldDarkRed{[}@*/ expression /*@\BoldDarkRed{]}@*/ | array-var /*@\BoldDarkRed{[}@*/ expression /*@\BoldDarkRed{]}@*/ /*@\BoldDarkRed{[}@*/ expression /*@\BoldDarkRed{]}@*/
expression: term | a-un-op expression | expression a-bin-op expression | boolean-expr /*@\color{darkred}{{\bf ?}}@*/ expression /*@\color{darkred}{{\bf :}}@*/ expression
\end{lstlisting}
&
\lstset{
   keywords=[1]{declaration,declaration-statement,program,list-of-statements,statement,conditional,loop,sync,assignment,target,boolean-expr,expression,return-statement,type,variable-decl,function-decl,property-decl,precondition,postcondition,property,quantifier,range,term,access,function-call,call-argument,argument-list,call-arg-list,var,array-var,array-declaration,then-block,else-block,while-block,target-term,array-access,fname,arg-decl-list,body-block,specname,pre-condition,post-condition,arg-call-list},
   keywords=[2]{@dotogether,@pre,@post,if,while,return,forall,exists,int,bool,num,else,constant,a-un-op,a-bin-op,b-un-op,b-bin-op,ba-bin-op,id},
}
\begin{lstlisting}
// statement extended with pre/post conditions
statement: assignment | conditional | loop | pre-condition | post-condition
// specification pre and postcondition 
pre-condition: @pre specname /*@\BoldDarkRed{\{}@*/ boolean-expr /*@\BoldDarkRed{\}}@*/
post-condition: @post specname /*@\BoldDarkRed{\{}@*/ boolean-expr /*@\BoldDarkRed{\}}@*/

// boolean expression extended with quantifiers
boolean-expr: term | b-un-op boolean-expr | boolean-expr b-bin-op boolean-expr | expression ba-bin-op expression | quantifier
quantifier: (forall|exists) /*@\BoldDarkRed{(}@*/ range /*@\BoldDarkRed{)}@*/ /*@\BoldDarkRed{\{}@*/ boolean-expr /*@\BoldDarkRed{\}}@*/   
range: var /*@\BoldDarkRed{[}@*/ expression /*@\BoldDarkRed{...}@*/ expression /*@\BoldDarkRed{]}@*/ 

// declaration statement extended with function declaration
declaration-statement: declaration /*@\BoldDarkRed{;}@*/ | function-decl 
function-decl: type fname /*@\BoldDarkRed{(}@*/ arg-decl-list?  /*@\BoldDarkRed{)}@*/ /*@\BoldDarkRed{\{}@*/ declaration-statement* body-block /*@\BoldDarkRed{\}}@*/
arg-decl-list: variable-decl ( /*@\BoldDarkRed{,}@*/ variable-decl)* 
body-block: list-of-statements return-statement
return-statement: return expression/*@\BoldDarkRed{;}@*/

// terms extended with function calls
term: constant | target-term | function-call 
function-call: fname /*@\BoldDarkRed{(}@*/ arg-call-list? /*@\BoldDarkRed{)}@*/
arg-call-list: expression ( /*@\BoldDarkRed{,}@*/ expression )*
specname, fname : id
\end{lstlisting}  
\\
\end{tabular}
\end{lrbox}

\begin{figure}[tb]
\resizebox{\textwidth}{!}{\usebox1}
\vspace{-2.0em}
\caption{Grammar of \psqcore (left), the core subset of the \psqlanguage imperative language (right).}
\label{fig:grammars}
%, the core subset and the full imperative  \psqcore, the core subset of the \psqlanguage~imperative language, (right) \psqlanguage with functions and specifications.}
\end{figure}

\subsection{\psqlanguage programs}

The grammar on the left of Figure~\ref{fig:grammars} defines \psqcore, 
the core subset of \psqlanguage. 
A \psqcore program is one or more declaration statements, 
followed by one or more statements. 
The directives \cci{a-un-op}, \cci{a-bin-op}, \cci{b-un-op}, 
\cci{b-bin-op}, and \cci{ba-bin-op} denote arithmetic unary and binary 
operators, Boolean unary and binary operators, 
and Boolean arithmetic operators, respectively. 
The variables matching the \cci{var} and \cci{array-var} rules in a program \Pm
form the sets of scalar variables $V=\set{v_1,v_2,\ldots,v_m}$ 
and array variables $A=\set{a_1,a_2,\ldots,a_n}$,
respectively.

\begin{definition}[terms]
  \rm A {\em target term} is either a variable $v\in V$, or 
 an array access term of the form $a[e]$ 
 which denotes the $e^{th}$ element of $a$ where $a\in A$ and 
 $e$ is an expression.
 A {\em term} is either a target term, or 
 a constant $c\in \mathbb{Z}$. 
\end{definition}

\begin{definition}[expressions]
  \rm An {\em expression} is a term, 
 an arithmetic expression of the form 
 $-e, e_1+e_2, e_1-e_2, e_1*e_2, e_1/e_2, e_1\%e_2$ 
 where $e,e_1,e_2$ are expressions and 
 $-,+,*,/,$ and $\%$ denote subtraction, 
 addition, multiplication, division and remainder,
 respectively. 
\end{definition}

\begin{definition}[Boolean expressions ]
  \rm A {\em Boolean expression} is either
  (1) a constant from the set $\mathbb{B}=\set{\true, \false}$, 
  (2) a binary Boolean arithmetic expression of the form 
  $e_1<e_2, e_1\le e2, e_1> e_2, e_1\ge e_2, e_1==e_2$ 
  where $e_1,e_2$ are expressions and $<,\le,>,\ge,$ and $==$ 
  denote smaller, less than or equal, bigger than, 
  bigger than or equal, and equal,
  respectively, 
  (3) a binary Boolean expression of the form $b\&\& b', b \| b', b \rightarrow b', b ==b'$,
  or (4) a unary Boolean expression of the form $\neg b$ where 
  $b,b'$ are Boolean expressions and 
  $\&\&, \|, !, \rightarrow ,$ and $==$ denote logical conjunction, disjunction,
  negation, implication, and equivalence, respectively.
\end{definition}

\begin{definition}[First order logic formula]
  \rm A {\em  first order formula} is either a Boolean expression,
  or a quantified formula of the form $Q q. b(q)$, where
  $Q\in \set {\forall, \exists}$ is a universal
  or an existential quantifier, $q$ is a 
  quantified variable, and $b(q)$ is a first order 
  formula with $q$ as a free variable. 
\end{definition}

\begin{definition}[Statement]
  \rm A {\em statement} is 
  (1) an {\em assignment} statement of the form 
  $t = e$ where $t$ is a target term and $e$ is an expression, 
  (2) a {\em list of statements} in a 
  well defined order that could be empty, 
  (3) a {\em conditional} statement 
  $\mathit{if}(b_1)\{~\mathit{then-block}\}$~$else~\{~\mathit{else-block}\}$ 
  where $b_1$ is a Boolean expression, and 
  $\mathit{then-block}$, and $\mathit{else-block}$ are lists 
  of statements,
  (4) a while loop statement 
  $\mathit{while}(b_2)\{ \mathit{while-block}\}$
  where $b_2$ is a Boolean expression and 
  $\mathit{while-block}$ is a list of statements,
  or (5) a declaration statement of the form 
  $type~v;$ or $type~a[constant];$ where $type$ is 
  either \cci{bool} or \cci{int} denoting the domain of the variable values, 
  $v$ and $a$ denote the variables in sets $V$ and $A$,
  and \cci{constant} denotes the size of the corresponding array. 
\end{definition}

%\begin{Verbatim}[fontsize=\relsize{-1.0}, numbersep=4pt,numbers=left]

The grammar of Figure~\ref{fig:grammars} (right) extends \psqcore 
%with function declarations, function calls, preconditions, and postconditions
to form \psqlanguage.
Boolean expressions are extended with existential and universal quantifiers that
quantify a Boolean expression over a range
of variable values defined by a start and an end expression. 
Statements are extended with pre- and postconditions that share the same
specification name (\cci{specname}). 
Declaration statements are extended with function declarations that take an argument 
declaration list, and a body block including a return statement. 
Terms are extended with function calls with an argument call list. 

\subsubsection*{\psqlanguage program semantics}

The semantics of a program is defined in terms of traces of
variable values across execution time. 
The assignment statements define the data flow of the program and their semantics
is defined in terms of updating the value of the target term with the value of
the expression of the assignment statement.
The lists of statements, conditional statements, loop statements, function calls, 
and return statements define the control flow of the program. 
The program starts execution at the first statement in its list of statements. 
%with the value of the expression, incrementing the execution time, 
%and moving execution control to the next statement.
The list of statements defines the order of execution. 
An if/else conditional statement $s$ in a list of statements $\ell$
executes the \cci{then-block} if the 
value of the Boolean expression $b$ is \true, otherwise it executes
the \cci{else-block}. 
The last statement in the \cci{then-block} and the \cci{else-block} 
moves execution to the statement next to $s$ in $\ell$. 
A loop statement $s$ in a list of statements $\ell$ 
executes the first statement of the \cci{while-block} if the 
value of the Boolean expression is \true, otherwise it executes the 
statement next to $s$ in $\ell$. The last statement in the \cci{while-block}
moves execution back to the loop statement $s$. 

A function call updates the values of the function argument declaration 
variables with the corresponding function call argument expressions and 
moves execution to the first statement of the \cci{body-block}. 
The return statement in a function declaration 
substitutes the corresponding function call with the value of the return 
expression and moves execution back to the statement that 
made the function call.

A specification \pair{\Pre}{\Post} is a pair of first order 
formulae where \Pre is the precondition specifying constraints
on the inputs of \Pm and \Post is a postcondition relating the
outputs of \Pm to its inputs. 

\subsection{AIG Sequential Circuit}
\label{sec:sequential}
%The ABC synthesis and verification framework reasons about And-Inverter-Graph 
%sequential circuits.

\begin{definition}[Sequential circuit]
\rm A {\em sequential circuit} is a tuple $\big( (U, E),G,
O\big)$.  The pair $(U,E)$ is a directed graph on
vertices $U$ and edges $E \subseteq U\times U$ where $E$
is a totally ordered relation.  The function $G: U \mapsto
{\mathit types}$ maps vertices to ${\mathit types}$.
There are three disjoint types: {\em primary inputs}, {\em
bit-registers} (which we often simply refer to as {\em
registers}), and logical {\em gates}.  Registers have designated
{\em initial value}, as well as {\em next-state
functions}.  Gates describe logical functions such as
the conjunction or disjunction of other vertices. 
A subset $O$ of $U$ specifies the {\em
primary outputs}.  
We denote the set of primary input variables by $I$,
and the set of bit-register variables by $R$.  
\label{def:back:seq_circuit}
\end{definition}

\begin{definition}[Fanins]
\rm We define the direct {\em fanin}s of a gate $u$ to be
$\{v \mid (v,u)\in E\}$ the set of source vertices connected
to $u$ in $E$.  We call the {\em support} of $u$ $\{v \mid
(v\in I \vee v \in R) \wedge (v,u) \in \ast E\}$ all
source vertices in $R$ or $I$ that are connected to $u$
with $\ast E$, the transitive closure of $E$.
\label{def:back:fanins} 
\end{definition}

%We restrict gates to have 2~fanins, and
%compute the NAND function; since NAND is functionally
%complete, this is not a limitation.  

For the sequential
circuit to be syntactically well-formed, vertices in $I$
should have no fanins, vertices in $R$ should have
2~fanins (the next-state function and the initial-value
function of that register), %gates should have two fanins,
and every cycle in the sequential circuit should contain
at least one vertex from $R$.  The initial-value functions
of $R$ shall have no registers in their support.  All
sequential circuits we consider will be well-formed.  

{\em The ABC analyzer reasons about AIG sequential circuits which are
sequential circuits with only NAND gates restricted to have 2~fanins.}
Since NAND is functionally complete, this is not a limitation.  

\subsubsection*{Semantics of sequential circuits}
\label{s:back:crct_semantics}

\begin{definition}[State]
\rm A {\em state} is a Boolean valuation to vertices in $R$. 
%A {\em concrete input} is a Boolean valuation to vertices in $I$.
\end{definition}

\begin{definition}[Trace]
\rm A {\em trace} is a mapping $t: U \times \mathbb{N} \rightarrow
\mathbb{B}$ that assigns a valuation to all vertices in
$U$ across time {\em steps} denoted as indexes from
$\mathbb{N}$.  The mapping must be consistent with $E$ and
$G$ as follows.  Term $u_{j}$ denotes the source vertex of
the $j$-th incoming edge to $v$, implying that
$(u_{j},v)\in E$.  The value of gate $v$ at time $i$ in
trace $t$ is denoted by $t(v,i)$.
\[
t(v,i)=
   \begin{cases}
      s^i_{v}            &:v \in I \ \text{with sampled value $s_{v}^i$}\\
      t(u_1, 0)       &:v \in R,i=0,u_1:=\ \text{initial-state of $v$}\\
      t(u_2, i-1)        &:v \in R,i>0,u_2:=\ \text{next-state of $v$}\\
      G_v\big(t(u_{1},i),...,t(u_{n},i)\big) &: v \ \text{is a combinational gate with function 
$G_v$}
   \end{cases} \newline
\]
\end{definition}

The semantics of a sequential circuit are defined with
respect to semantic traces.  Given an input valuation
sequence and an initial state, the resulting trace is a
sequence of Boolean valuations to all vertices in $U$
which is consistent with the Boolean functions of the
gates.  We will refer to the transition from one valuation
to the next as a {\em step}.  A node in the circuit is
satisfiable %justifiable 
if there is an input sequence which when
applied to an initial state will result in that node
taking value \true.  A node in the circuit is
valid if its negation is not satisfiable.  We will refer
to targets and invariants in the circuit; these are 
vertices whose satisfiability and validity
is of interest, respectively.
A sequential circuit can naturally
be associated with a finite state machine (FSM),
which is a graph on the states.  However, the 
circuit is different from its FSM; among
other differences, it is exponentially more succinct in
almost all cases of interest~\cite{BuClMcDiHw92}. 

%Figure~\ref{f:back:cctfsm}(a) shows a 2-bit write enabled 
%buffer which uses a multiplexer enabled by $w_{en}$ to 
%update its register bits $r$ with values in inputs $i$ 
%when $w_{en}$ is $1$ and retains its state otherwise.
%It also shows the corresponding FSM with no labels on
%transitions for clarity of exposition purposes.
%Figure~\ref{f:back:cctfsm}(b) shows a 3-bit version of 
%the same circuit with the corresponding FSM. 
%Only the arcs corresponding to state $000$ are shown for 
%clarity of exposition.

\subsection{ABC synthesis and verification framework}
\label{s:back:abc}
\begin{table*}[bt]
\centering
\caption{Brief description of selected ABC synthesis and verification techniques.}
\vspace{-1em}
\resizebox{.9\textwidth}{!}{
\begin{tabular} {p{2.4cm}p{13cm}l}
\hline
\centering{{\bf Technique}} & {\bf Description} & {\bf Command} \\
\hline
Balancing ~\cite{brayton2010abc} & Logic balancing applies associativity transformation to reduce AIG levels. & 
\cci{balance} \\

Sweep & Structural register sweep (SRS) reduces the number of registers in the circuit
by eliminating stuck-at-constant registers~\cite{mishchenko2008scalable}.&
\cci{ssweep} \\

Correspondence & Signal correspondence (Scorr) computes a set of classes of 
sequentially-equivalent nodes using $k$-step induction~\cite{mishchenko2008scalable}.& 
\cci{scl -l} \\

Rewriting & AIG rewriting iteratively selects and replaces 
rooted subgraphs with smaller pre-computed subgraphs in order to reduce AIG size~\cite{bjesse2004dag}.& 
\cci{rewrite} \\

Refactoring & Refactoring is a variation of rewriting. It uses a heuristic
to compute a large cut for selected AIG nodes, then replaces the sub-graph that 
corresponds to the cut with a refactored structure
if an improvement is observed~\cite{mishchenko2006dag}.& 
\cci{refactor}  \\

Retiming & Retiming manipulates register boundaries and count in a given logic network, while maintaining output 
functionality and logic structure~\cite{hurst2007fast}.& 
\cci{retime}\\

%-- Verification -- %
Induction & Temporal induction uses circuit SAT and BDD solvers
to carry simple and k-step induction proofs over the time steps of the AIG~\cite{een2003temporal}.& 
\cci{ind} \\

Interpolation & Interpolation-based algorithms aim find interpolants and 
overapproximate the reachable states
of the AIG with respect to the property~\cite{amla2005analysis}.& 
\cci{int} \\

Reachability & Property directed reachability (Pdr) tries to prove that 
there is no transition from an initial state of the AIG to a bad state~\cite{een2011efficient}. & \cci{pdr} \\
\hline
\end{tabular}
}
\normalsize
\label{t:back:transforms}
\vspace{-2.5em}
\end{table*}

% -- Synthesis techniques -- %
ABC is an open source synthesis and verification framework for sequential circuits.
ABC operates on sequential circuits in AIG format and checks the satisfiability 
of a designated output gate therein. 
ABC applies several reduction and abstraction
techniques to simplify and decompose the problem into smaller problems. 
It then calls decision techniques to decide the simplified problems. 
Table~\ref{t:back:transforms} briefly summarizes some of the techniques supported
by ABC. 

\comment{
In what follows we discuss some of the techniques that are briefly 
list in Table~\ref{t:back:transforms}. 

%\subsubsection {Structural Register sweep \cci{ssweep}}
\subsubsection{Structural Register Sweep (SRS)}
SRS detects registers that are stuck-at-constant and eliminates 
them from a given sequential AIG circuit. The technique starts by zeroing up all 
initial values of registers in the circuit. It then uses the ternary simulation
algorithm in order to detect stuck-at-constant registers. The algorithm starts from 
the initial values of the registers and simulates the circuit using x values for the
circuit's primary inputs. The simulation algorithm stops when a new ternary state is 
equal to a previously computed ternary state. In this case, any register having the 
same constant value at each reachable ternary state will be declared to be 
stuck-at-constant and thus eliminated. The structural sweeping algorithm stop when 
no further reduction in the number of registers is possible~\cite{mishchenko2008scalable}. 

%\subsubsection {Signal correspondence \cci{scorr}}
\subsubsection{Signal Correspondence (Scorr)} 
Scorr uses $k$-step induction in order to detect and merge sets of classes of 
sequentially-equivalent nodes~\cite{mishchenko2008scalable}. The base case for this algorithm is that the equivalence
between the classes holds for the first $k$ frames, and the inductive case is that 
given the base case, starting from any state, the equivalence holds in the 
$(k+1)^{st}$ state. Key to the signal correspondence algorithm is the way the candidate
equivalences are assumed for the base case. Abc implements speculative reduction, 
originally presented in~\cite{mony2005exploiting}, which merges, but does not remove, any node of an equivalence 
class onto its representative, in each of the first $k$ time frames. Instead of removing the 
merged node, a constraint is added to assert that the node and its representative are equal. 
This technique is claimed to decrease the number of constraints added to the SAT solved for 
induction. 

\subsubsection{Rewriting}
Rewriting aims at finding nodes in a Directed Acyclic Graph (DAG) where by replacing subgraphs rooted 
at these nodes by pre-computed subgraphs can introduce important reductions in the DAG size, while 
keeping the functionality of these nodes intact. The algorithm traverses the DAG in depth-first post-order
and gives a score for each root node. The score represents the number of nodes that would result
from performing a rewrite at this node. If a rewrite exists such that the size of the DAG is decreased, such 
a rewrite is performed and scores are recomputed accordingly.  
Rewriting has been proposed initially in~\cite{bjesse2004dag}, targeted for Reduced Boolean Circuits (RBC); 
it was later implemented and improved for ABC in~\cite{mishchenko2006dag}. 

% Retiming
\subsubsection{Retiming}
Retiming a sequential circuit is a standard technique used in sequential synthesis, 
aiming at the relocation of the registers in the circuit in order to optimize 
some of the circuit characteristics. Retiming can either targets the minimization of the delay 
in the circuit, or the minimization of the number of registers given a delay constraint, 
or the unconstrained minimization of the number of registers in the circuit. It 
does so while keeping the output functionality of the circuit intact~\cite{hurst2007fast}

% -- Verification techniques -- %
%\subsubsection {Property directed reachability \cci{pdr}}
\subsubsection{Property Directed Reachability (Pdr)}
The Pdr algorithm aims at proving that no 
violating state is reachable from the initial state of a given AIG network. 
It maintains a trace representing a list of over-approximations of the states
reachable from the initial state, along with a set of {\em proof-obligations}, 
which can be a set of bad states or a set of states from which a bad state is
reachable. Given the trace and the set of obligations, the Pdr algorithm manipulates 
them and keeps on adding facts to the trace until either an inductive invariant 
is reached and the property is proved, or a counter example is found (a bad is state
is proven to be reachable). The algorithm was originally developed by Aaron Bradley 
in~\cite{bradley2011sat, bradley2007checking} and was later improved by Een et. al in~\cite{een2011efficient}.

\subsubsection{Temporal Induction}
Temporal induction carries an inductive proof of the property 
over the time steps of a sequential circuit.
Similar to a standard inductive proof, it consists of a base
case and an inductive hypothesis. These steps are typically 
expressed as SAT problems to be solved by traditional SAT solvers.  
$k$-step induction strengthens simple temporal inductive proofs 
by assuming that the property holds for the first $k$ time steps (states), 
i.e. a longer base case needs to be proven~\cite{een2003temporal}. Since the target is
to prove unsatisfiability (proving that the negation of the property 
is unsatisfiable), if the base case is satisfiable, a counter-example 
is returned. Otherwise, the induction step is checked by assuming that
the property holds for all the states except the last one (the $(k+1)$'th 
state)~\cite{biere2009handbook}.   

\subsubsection{Interpolation}
Given an unsatisfiable formula $A \land B$, an interpolant $I$ is
a formula such that $A \implies I$, $I \land B$ is unsatisfiable and
$I$ contains only common variables to $A$ and $B$. 
Given a system $M$, a property $p$ and a bound $k$, interpolation
based verification starts by attempting bounded model checking (BMC) with the bound $k$. 
If a counter-example is found, the algorithm returns. Otherwise, it
partitions the problem into a prefix $pre$ and a suffix $suf$, such that the 
problem is the conjunction of the two. 
Then the interpolant $I$ of $pre$ and $suf$ is computed, it represents
an over-approximation of the set of states reachable in one step from the initial state
of the algorithm. If $I$ contains no new states, a fixpoint is reached 
and the property is proved. Otherwise, the algorithm reiterates and replaces
the initial states with new states added by $I$~\cite{amla2005analysis}. 
}

\section{Overview}
\label{s:overview}

\comment{Given an imperative program \Pm, 
a precondition and postcondition FOL specification pair \pair{\Pre}{\Post}, 
and a bound $b$ on the domain of \Pm~and its variables,
\psqtool ~checks whether \Pm satisfies its specifications 
($\Pm \models \pair{\Pre}{\Post}|_{b}$); i.e. when the bounded inputs 
of \Pm satisfy \Pre, the outputs of \Pm must satisfy \Post.
\psqtool translates the check into an AIG circuit with
a designated output $o$ that is set to \true iff 
\Pm violates its specifications. 
}

\begin{figure}[bt]
 \centering
 \resizebox{\textwidth}{!} {
     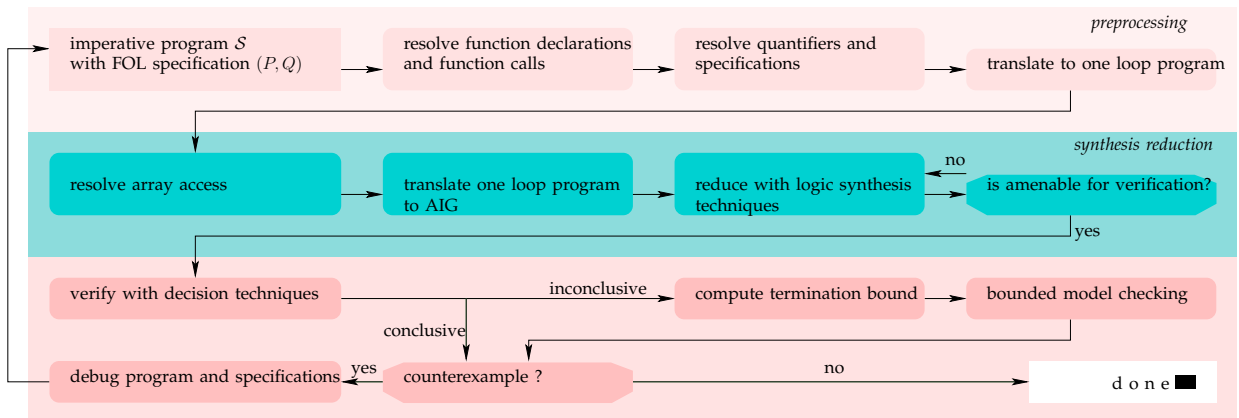
 }
 \caption{Overview of verifying \psqlanguage programs with AIG synthesis and verification techniques}
   \label{f:flow}
   \vspace{-2em}
\end{figure}

Figure~\ref{f:flow} illustrates our method. 
First, 
\psqtool preprocesses program \Pm with its FOL specification pair \pair{\Pre}{\Post} 
to resolve function calls, specifications, and quantifiers and 
produces an equivalent program \Pm' in \psqcore. 
%as written in \psqlanguage program with specifications as described in the 
%grammar of Figure~\ref{fig:fullgrammar} and resolves function calls and quantifiers to 
%produce an equivalent \psqcore program. 
%restricted to the core \psqlanguage grammar of Figure~\ref{fig:grammar}.
It then translates \Pm' into a one loop program where it encodes
the control flow into an additional program counter variable and encodes the data 
flow into assignment statements with ternary conditional expressions that depend
on the value of the program counter variable. 
Then \psqtool translates the generated one loop program into an AIG circuit
with bit vectors of fixed width that correspond to program variables. 
%The translation instantiates a vector of AIG registers to correspond to each 
%program variable. 
Array sizes are limited by the largest possible index %$2^b-1$ 
and \psqtool translates each array element into a vector of AIG registers. 
\psqtool resolves array access operations to refer to the vectors of registers, 
it then instantiates equivalent logic circuits to the expressions and 
connects the circuits to the initial value and next state value functions 
of the registers.
\psqtool designates one output of the AIG circuit to be \true when the program
violates the specifications. 
Other AIG outputs signal out of bound array access and arithmetic overflow.

\psqtool uses the ABC synthesis techniques to reduce the size and 
the complexity of the generated AIG circuit until it is amenable for verification. 
Once the AIG circuit is small enough, \psqtool uses ABC verification algorithms
to check the designated output. 
The verification technique may find a violation and return a counterexample, 
return a proof that the program is correct within the bound, or return
an inconclusive result after exhausting computational limits. 
When an ABC verification returns a counterexample,
\psqtool translates the counterexample from the AIG circuit level to traces
of variable values in the original program and provides the user with a visual 
debugging view using GTKWave~\cite{gtkwave}.

%In case all verification techniques returned inconclusive results, 
%then \psqtool uses heuristics to compute a guess on the number of execution time
%steps needed for the program to terminate (the program counter variable in the
%AIG circuit reaches its last label value). 
%The \psqtool uses bounded model checking to check whether the termination 
%bound is correct and whether the program violates the specification within 
%the termination bound. 

We describe the termination guarantee bound technique, the \oneloop programs, relate them to 
AIG circuits, and then illustrate the translation to \oneloop
using an array search example. 

\subsection{Termination guarantee bound}

In case the ABC prove techniques do not reach a conclusive result, 
\psqtool uses heuristics to compute a termination upper bound $\theta$ in terms
of execution time, and calls ABC twice. 
The first call is to verify that the program 
is guaranteed to terminate within the termination bound 
($\psi\equiv time >= \theta \rightarrow pc = last(\Pm)$). 
The second call uses $\theta$ as a bound with the ABC bounded model checking 
technique to prove that the program does not violate the specification within 
the termination bound. 
Intuitively, checking 
$\Pm \sat \pair{\Pre}{\Psi}|_b$ is often easier than checking \prob,
and once $\theta$ is confirmed as an upper bound for termination, 
bounded model checking is decidable. 

\subsection{One loop programs}
\label{s:oneloop}
\lstset{
   keywords=[1]{primary-input,
   primary-declaration,declaration,declaration-statement,
 decl-list,init-list,next-list,one-loop-program,assignment},
   keywords=[2]{@dotogether,@pre,@post,if,while,return,forall,exists,int,bool,num,else,constant,a-un-op,a-bin-op,b-un-op,b-bin-op,ba-bin-op,notdone,primary},
}
%\begin{figure}
  \begin{wrapfigure}{R}{0.36\textwidth}
%  \centering
%  \begin{minipage}{6cm}
\begin{lstlisting}
primary-input: primary declaration/*@\BoldDarkRed{;}@*/

decl-list: primary-input* declaration-statement+ 
init-list: assignment+
next-list: assignment+

one-loop-program: decl-list 
    @dotogether /*@\BoldDarkRed{\{}@*/
      init-list /*@\BoldDarkRed{\}}@*/

    while /*@\color{darkred}{{\bf (}}@*/notdone/*@\color{darkred}{{\bf )}}@*/ /*@\color{darkred}{{\bf \{}}@*/ @dotogether /*@\color{darkred}{{{\bf \{}@*/
      next-list /*@\color{darkred}{{\bf \}}}@*/ /*@\color{darkred}{{\bf \}}}@*/
\end{lstlisting}
%\end{minipage}
\vspace{-1.5em}
\caption{\oneloop Grammar}
\label{fig:oneloopgr}
%\end{figure}
\end{wrapfigure}

The grammar in Figure~\ref{fig:oneloopgr} extends \psqcore with
the $primary$ and $dotogether$ constructs to define 
one loop programs.
The \cci{primary} construct denotes
a primary input variable declaration with non-deterministic values. 
The concurrent $dotogether\{together\textit{-}block\}$
denotes that all statements of the $together\textit{-}block$ list of statements execute
simultaneously. 
%
%If the list of statements is directly within a $\mathit{dotogether}$
%statement, then all statements execute at once. 
%
An \oneloop starts with variable declarations.
Then the \cci{init-list} list of assignment statements concurrently initializes 
the \oneloop variables. 
After initialization, a loop keeps updating the values of the variables
concurrently using the list of assignment statements \cci{next-list}
until the \oneloop is done as denoted by the $notdone$ Boolean variable. 
Intuitively, the structure of an \oneloop is similar to that of a sequential
circuit where the variables correspond to registers, the \cci{init-list}
and \cci{next-list} correspond to the initial and next state value functions,
respectively.

%All the assignment statements within \cci{next-list} execute simultaneously as indicated with the \cci{do-together} keyword.

%Each assignment statement has a left hand side target term 
%which is either a variable or an access operator to an 
%array element. 
%The right hand side of an assignment is a combinational expression ranging over the program variables,  Boolean and arithmetic operators, and a ternary choice 
%operator. The ternary choice \cci{(a? b : c)} returns $b$ if $a$ 
%is \true and $c$ otherwise. 

\subsection{Array search example }
\label{s:arrayexample}
\lstset{
  keywords=[1]{for,int,while,if,else,return,@pre,@post,forall,break,xor,false,true,@dotogether},
  keywords=[2]{a,d,s,e,n,MAXSIZE,i,rv,as,preas,postas,pc,notdone},
  keywords=[3]{0,1,2,3,4,5,6,7,8,9,10,11,12,13,14,15,16,17,18,19,20,-1},
}
\lstset{ firstnumber=1,numbers=left, numberstyle=\tiny, stepnumber=1, numbersep=5pt}

\begin{lrbox}{0}
\begin{tabular}{p{3.4in}cp{3.5in}}
\begin{lstlisting}
int /*@\textbf{ArraySearch}@*/(int[] a,int d,int s,int e,int n) {
 @pre as { 0<=s && s<=e && e<n && n<=MAXSIZE; }
 int i = s;         // pc = pc+1
 while ( i <= e) {  // pc = (i <= e) ? 5 : 10;
  if (a[i] == d) {  // pc = (a[i] == d) ? 6 : 8;
   break;}          // pc = 10;
  else {
   i = i+1;}        // pc = 4;
  }
  return i;         // pc = 11; rv = i;
@post as {
 (s <= rv && rv <= e) -> a[rv] == d
 forall(int i)[s .. e]{
   a[i] != d -> (rv== -1) }
} }
\end{lstlisting}
& & 
%\begin{Verbatim}[fontsize=\relsize{-2}], numbersep=4pt,numbers=left]
\begin{lstlisting}
@dotogether { // init-list for initial value0s
 preas = 0 <= s && s <= e && e < n && n <= MAXSIZE;
 pc = 0; notdone = true; postas = true; }
while (notdone) { @dotogether {   
 // next-list with next state functions
 i = (pc == 3) ? s : (pc == 8) ? i+1 : i;
 notdone = (pc == 11) ? false : true;
 rv = (pc == 10) ? i : rv;
 pc = (pc == 0) ? 3: (pc == 3) ? 4 : 
      (pc == 4) ? ( (i <= e) ? 5 : 10 ) :
      (pc == 5) ? (a[i] == d) ? 6 : 8 ) :
      (pc == 6) ? 10 : (pc == 8) ? 4  : 
      (pc == 10) ? 11 : pc; 
 postas = (rv >= s && rv <= e) -> a[rv] == d &&
  forall(int i)[s .. e]{a[i] != d -> (rv== -1); }}}
\end{lstlisting}
\end{tabular}
\end{lrbox}

\begin{figure}[tb]
  \resizebox{1.01\textwidth}{!}{\usebox0} 
\vspace{-2.5em}
\caption{Array search in \psqlanguage and its equivalent \oneloop with the program counter variable.}
\label{fig:arraysearch}
\vspace{-2.5em}
\end{figure}
\lstset{numbers=none}

The Array search program in Figure~\ref{fig:arraysearch}
takes as input an array $a$, 
a start index $s$, an end index $e$, a data value $d$, and the number of 
elements in the array $n$.
%It is annotated with a specification consisting of a precondition and a postcondition. 
The precondition states that $s$ and $e$ are within array
bounds and that $n$ is within the bound of array sizes. 
The postcondition makes use of a variable $rv$ denoting the return value. 
It states that if $rv$ is valid between $s$ and $e$ inclusive, 
then $a[rv]$ must be equal to $d$, 
otherwise, $rv$ must be invalid (-1) and all entries in $a$ between $s$ 
and $e$ inclusive are not equal to $d$. 
Figure~\ref{fig:arraysearch} also shows a semantically equivalent \oneloop array search 
program in terms of the values of the common program variables at termination.

The equivalent program introduces Boolean variables \cci{preas}, \cci{postas}, and 
\cci{notdone} to encode the precondition, postcondition, and termination
state of the program, respectively. 
The equivalent program also introduces a program counter variable
\cci{pc} which encodes the control flow of the program as indicated in 
the comments of the original program.% in Figure~\ref{fig:arraysearch}. 
Variable \cci{notdone} is initialized to $true$, and \cci{pc} 
is initialized to the first executable line of the program $3$. 
Once \cci{pc} reaches the last executable line of the program $13$, 
the program terminates and thus \cci{notdone} becomes $false$. 
Assignment statements are grouped by target variables, and encoded
into ternary conditional expressions that depend on the value of \cci{pc}. 
For example, the iterator $i$ is assigned to $s$ when \cci{pc} is $3$, incremented
when \cci{pc} is $8$, and remains the same otherwise. 
Furthermore, the assignment statements on Lines 2 and 3 assign initial values to the 
target variables. 
The assignment statements inside the \cci{while} loop (Lines 5 to 15) compute 
the next state value of each of the target variables. 

Our method translates the \oneloop program to an AIG 
sequential circuit where an iteration of the \cci{while} loop
is equivalent to a single time step in the AIG. 
%
%The method represents each Boolean variable with one register, and each scalar 
%variable with a finite vector (bit-vector) of registers 
%with initial value and next state functions. 
%The initial state functions of the vector of registers corresponding to
%a variable are connected to a vector of gates that represents the right hand 
%side initial value assignment statement of the variable. 
For example, 
The \cci{pc} variable ranges from $0$ to $11$ and is encoded with bit-vector of
$4$ registers
with initial value functions set to $0$ and next state functions set to gates
representing the right hand side expression of the $pc$ assignment statement. 
%The sequential circuit in Figure~\ref{f:pccircuit} shows the registers,
%their initial value functions set to $0$, and their next state functions
%set to gates reprsenting multiplexers that chose the next value of the
%\cci{pc} registers based on the current state of the cuicuit. 
%
%Program variables that are not initialized in the code are considered
%input variables and the methods connects their initial value functions 
%to free primary inputs.
%The method connects the next state functions of register vectors corresponding to 
%program variables to gates that represent
%the right hand side of the next state assignment.
%The conditional, arithmetic, and Boolean operations in the right hand side
%expressions are encoded as combinational logic circuits in the usual manner. 
Section~\ref{s:onelooptoaig} discusses the translation from \oneloop to AIG. 

Our method takes the resulting AIG and designates a gate therein that 
represents $\neg(preas \wedge done \rightarrow postas)$ as the output gate, passes
the AIG to ABC, and checks for validity.
The ABC solver returns the counterexample of Figure~\ref{fig:vdebug} 
with $a=[15~15~15~11],s=3,e=3,n=4,d=13,rv=0$ 
where $d$ is not in $a$, and the return value is $e+1$, while the postcondition
requires an invalid index $(-1)$. 
The provided counter example can be used to fix the program. 
A possible fix is to replace Line 6 with \cci{return i;}, and Line 10 with
\cci{return -1;}.
Our method takes the fixed program, translates it into a sequential circuit,
and passes it to ABC which validates the correctness of the fixed program
%modulo the 
%finite size of the domain of the variables. variable vectors 
using symbolic model checking.

%%%%%%%%%%%%%%%%%%%%%%%%%%%%%%%%%%%%%%%%%%%%%%%%%%%%%%
%%%%%%%%%%%%%%%%%%%%%%%%%%%%%%%%%%%%%%%%%%%%%%%%%%%%%%

\section{Transformation}
\label{s:transformation}
In this section we present source to source transformation 
algorithms to translate a program \Pm written in \psqcore  
into a \oneloop program.
We then present algorithms that preprocess a \psqlanguage program with 
functions and specifications and translates it into \psqcore. 
Finally, we present algorithms for translating the generated \oneloop program
into an AIG circuit.
The algorithms follow the execution semantics of \psqlanguage
%, the \cci{dotogether} construct 
and the correctness of the translation holds by construction. 
We first present helper functions and terms that are used in the 
algorithms. 
\subsection{Helper functions} 
\label{s:helper}
Function \cci{label($s$)} takes a statement in \Pm and returns a unique label 
in $\mathbb{N}$. 
Function \cci{next($s$)} takes a statement $s$ and returns the label of the next 
statement of $s$ in a list of statements $\ell$. 
If $s$ is the last statement in $\ell$, several cases arise.
\begin{enumerate}
\item $\ell$ is a function declaration \cci{body-block}, \cci{next} returns the label of the return statement.
\item $\ell$ is a \cci{while-block} of a loop $l$, \cci{next} returns the label of the loop statement \cci{label($l$)}.
\item $\ell$ is a \cci{then-block} or an \cci{else-block} of a conditional $s_1$, 
  \cci{next} returns the label of the statement next to $s_1$ (\cci{next($s_1$)}).
\item Finally, $\ell$ is the last statement in \Pm, \cci{next} returns a 
  special label $done$. 
\end{enumerate}
Functions \cci{first} and \cci{last} return the first and the last 
statement in a list of statements, respectively, or $nil$ if the list is empty. 
Functions  \cci{then}, and \cci{else} take 
a conditional statement $s$ of the form 
\cci{if ($b$) \{then-block\}} \cci{else \{else-block\}}.
%where $b$ is a
%Boolean expression, 
%\cci{then-block} and \cci{else-block}
%are lists of statements. 
They return the labels of the first statement of the 
then and else blocks, 
\cci{label(} \cci{first( } \cci{then-block))}, and \cci{label(} \cci{first( } \cci{else-block))},
respectively.
If the list of statements is empty, both functions 
return \cci{next($s$)}. 

Function \cci{body} takes a loop statement or a function declaration. 
The loop statement is of the form
\cci{ while ($b$) \{while-block\}}.
%where $b$ is a Boolean expression and \cci{while-block} is a list of statements.
The function declaration is of the form \cci{type} \cci{fname(} \cci{arg-decl-list)} \cci{ \{ body-block \} }
where \cci{type}, \cci{fname}, and \cci{arg-decl-list} are the return type, 
name, and argument declaration list of the function, respectively, and
\cci{body-block} is a list of statements followed by a return statement. 
For a function declaration, \cci{body} returns the label of the first statement 
in \cci{body-block} \big(\cci{label(first(body-block))}\big).
For a loop, \cci{body} returns the label of the first statement in \cci{while-block}. 
In case \cci{while-block} is empty, it returns the label of the loop. 
Function \cci{condition} takes a conditional or a loop statement 
and returns $b$. 

Functions \cci{target} and \cci{expr} take an assignment statement 
of the form \cci{target-term =} \cci{expression;}  and
return \cci{target-term} and \cci{expression}, respectively. 
Function \cci{base} and \cci{index} take 
an array access term of the form \cci{ $a$[$e_1$]} or 
\cci{$a$[$e_1$][$e_2$]} where $a$ is an array variable, and $e_1$ and $e_2$ are
expressions. 
Function \cci{base} returns $a$. 
When $a$ is declared as \cci{type $a$[$c_1$]}, where $c_1$ is a constant, 
\cci{index} returns $e_1$. 
When $a$ is declared as \cci{type $a$[$c_1$][$c_2$]}, where $c_2$ is a constant,
\cci{index} returns $e_1c_2 + e_2$. 
Finally, function \cci{typeof} takes a variable (or array variable) and returns its type.

\subsection{\psqcore programs to \oneloop}
\label{s:simple2olp}
\lstset{
   keywords=[1]{foreach,with,endfor,in,if,elseif,endif,to,else,let,be},
   keywords=[2]{append,next-list,v-next-list,v-expr,pc-next-list,decl-list,init-list,a-index,a-expr,a-next-statement},
}
\begin{figure}
\begin{tabular}{p{7.6cm}p{8cm}}
\begin{lstlisting}
/*@\textbf{generateOneLoopProgram(\Pm)}@*/
 /*@\textbf{generateDeclarationList($\Pm$)}@*/ 
 /*@\textbf{generareInitList($\Pm$)}@*/
 // translate data flow into next-list 
 foreach variable $v$ in $V\cup A$
  if $v \in V$
   v-next-list = /*@\textbf{generateVarNextList($v$,$\Pm$)}@*/
  else // $v \in A$
   v-next-list = /*@\textbf{generateArrayNextList($v$,$\Pm$)}@*/
  endif
  append v-next-list to next-list 
 endfor 
 // translate control flow into next-list
 pc-next-list = /*@\textbf{generatePCNextList($\Pm$)}@*/
 append pc-next-list to next-list
 append $notdone~=~!(pc == done);$ to next-list
\end{lstlisting}

\begin{lstlisting}
/*@\textbf{generateVarNextList($v$,\Pm)}@*/
 foreach statement $s$ with target($s$) = $v$ 
  append $(pc==$label$(s))~?~$expression(s)$~:$ to v-expr
 endfor
 //otherwise $\textcolor{darkgreen}{v}$ stays the same
 append $v;$ to v-expr 
 let v-next-list be $v = $ v-expr $;$
\end{lstlisting}

&  
\begin{lstlisting}
/*@\textbf{generateDeclarationList(\Pm)}@*/
 foreach variable $v$ in $V \cup A$
  append typeof$(v)~v;$ to decl-list
  // handle variables that are not assigned
  if $v$ is used before it is assigned 
   append $primary$ typeof$(v)~vnondet;$ to decl-list
  endif
 endfor 
 append $int~pc;$ 
        $bool~notdone;$ to decl-list
\end{lstlisting}

\begin{lstlisting}
/*@\textbf{generateInitList(\Pm)}@*/
 foreach variable $v $ in $V \cup A$
  //variables used before assignied are nondeterministic
  if $v$ is used before it is assigned 
   append $v=vnondet;$ to init-list
  else 
   append $v=0;$ to init-list
  endif
 endfor 

 // initialize program counter
 append $pc=$label(first$(\Pm));$ 
        $notdone=true;$ to init-list
\end{lstlisting}
\end{tabular}
\vspace{-2em}
\caption{Algorithms for generating the \oneloop \cci{decl-list}, \cci{init-list}, and \cci{next-list} from \Pm}
\label{f:decl}
\vspace{-2em}
\end{figure}

Algorithm \cci{generateOneLoopProgram} of Figure~\ref{f:decl} takes a 
program \Pm written in \psqcore and generates 
the \cci{decl-list}, \cci{init-list}, and \cci{next-list} 
of an equivalent \oneloop program. 
Algorithm \cci{generateDeclarationList} generates a declaration for each 
variable in \Pm, and also generates a primary input \cci{vnondet} variable
for each variable that is used before being initialized in \Pm. 
It also declares the program counter $pc$ and the termination 
guard $notdone$ variables. 
Algorithm \cci{generateOneLoopProgram} calls \cci{generateInitList}
that generates \cci{init-list} where all variables are initialized to $0$
except those that are used before being assigned in \Pm, 
those are initialized with their corresponding 
non-deterministic primary input variables. 
The program counter $pc$ and the termination guard $notdone$ are initialized
to the label of the first statement of the program, and to \true, respectively. 

The \cci{generateOneLoopProgram} Algorithm 
builds one assignment statement \cci{v-next-list}
for each variable $v$ and appends it to \cci{next-list}. 
If $v$ is a regular variable,
\cci{generateVarNextList} iterates over each assignment $s$ 
where $v$ is the target and builds a nested ternary conditional expression \cci{v-expr}
that evaluates to \cci{expr(s)} when $pc$ points to the label of $s$.
When $pc$ does not point to a statement that assigns to $v$, 
\cci{v-expr} evaluates to the original value of $v$ which is 
is denoted by appending $v$ as the value of the last choice in 
in the nested ternary expression.

%pc-next-list = /*@\textbf{generateVarNextList($pc$, $\Pm$)}@*/
\begin{figure}
\begin{tabular}{p{7.8cm}p{8.4cm}}
\begin{lstlisting}
/*@\textbf{generateArrayNextList($v$,\Pm)}@*/
 //iterate over all statements assigning to $\textcolor{darkgreen}{v}$
 foreach assignment $s$ with base(target($s$)) = $v$
  //$s$ is of the form $v[ie]=e$
  let $ie$ be index(target($s$))
  let $e$ be expr($s$)   
  //aggregate array access in one expression
  append $(pc == $label$(s))~?~ie~:$  to a-index
  append $(pc == $label$(s))~?~e~:$ to a-expr
 endfor
 //when no array access happens, $\textcolor{darkgreen}{v[0]}$ does not change
 append $0$ to a-index
 append $v[0]$ to a-expr
 //construct one assignment statement for the array
 let v-next-list be $v[$a-index$] = $a-expr$;$
\end{lstlisting}
\caption{Algorithm to aggregate array variable assignments (left) and 
encode control flow into program counter $pc$ (right)}
\label{f:arrayassign}
\label{f:pc}
& 
\begin{lstlisting}
/*@\textbf{generatePCNextList(\Pm)}@*/
 // encode control flow into $\textcolor{darkgreen}{pc}$
 let pc-next-list be $pc = $
 // iterate over all statements
 foreach $s$ in $\Pm$
  if $s$ is a conditional  //handle conditional
   append $(pc == $label$(s))~?$ 
            $($condition$(s)~?~$then$(s)~:$ $\mathit{else}(s))~:$  
          to pc-next-list
  elseif $s$ is a loop     // handle loops
   append $(pc == $label$(s))~?$
            $($condition$(s)~?~$body$(s)~:$ next$(s))~:$ 
          to pc-next-list
  elseif $s$ is an assignment with target($s$) $=pc$
   // handle direct assignment to $\textcolor{darkgreen}{pc}$
   // which result from function call resolution 
   append $(pc == $label$(s))~?~$expr$(s) :$ to pc-next-list
  else // handle other statements (assignments)
   append $(pc == $label$(s))~?~$next$(s) :$ to pc-next-list
  endif
 endfor
 // when $\textcolor{darkgreen}{done}$, program counter does not change
 append $pc;$ to pc-next-list
\end{lstlisting}
\end{tabular}
\vspace{-3em}
\end{figure}

If $v$ is an array variable, Algorithm \cci{generateArrayNextList} 
from Figure~\ref{f:arrayassign} iterates over all statements $s$ where 
$v$ is target. It aggregates all expressions and index expressions into two
nested ternary conditional expressions \cci{a-index} and \cci{a-expr} 
for the index expression and the right hand side expression of the \cci{v-next-list}
assignment statement. 
The ternary conditional expressions depend on the position of $pc$ to return the 
corresponding expressions and index expressions. 
In case $pc$ does not point to a statement with $v$ as a target, then \cci{v-next-list}
makes sure that $v[0]$ stays the same by appending $0$ and $v[0]$ as the last
default choices in the ternary conditional expressions, respectively. 

Finally, \cci{generateOneLoopProgram} calls \cci{generatePCNextList} 
from Figure~\ref{f:pc} to encode 
the control flow of \Pm into \cci{pc-next-list} with
a nested ternary conditional expression that defines the value of $pc$. 
Algorithm \cci{generatePCNextList} iterates over all statements and encodes
their execution flow semantics. 
When $s$ is at a conditional statement, $pc$ moves to \cci{then-block} 
if the Boolean condition evaluates to \true, and to the \cci{else-block} otherwise. 
When $s$ is at a loop statement, then $pc$ moves to the first statement of 
\cci{body-block} if the Boolean condition evaluates to \true, and to the 
statement next to the loop otherwise. 
When $s$ is an assignment statement such that \cci{target($s$)}$=pc$, then 
$pc$ moves to the expression of $s$; this is similar to a \cci{goto} statement and
may result from resolving function calls as discussed
in Section~\ref{s:function}.
For other statements, $pc$ points to the next statement.

\subsection{Preprocessing \psqlanguage programs}
\label{s:psqtospq}
\psqtool~resolves function declarations, return statements, 
functions calls, pre and post conditions, and quantifiers
by translating \Pm into \Pm' that is restricted to \psqcore. 
%the core subset 
%of \psqlanguage. 

\lstset{
   keywords=[1]{foreach,with,endfor,in,if,elseif,endif,to,else,let,be,by,
   substitute,replace},
   keywords=[2]{append,prepend,
     next-list,decl-list,init-list, statement-list,
   type,fname,arg-decl-list,func-decl-list,func-body,func-return-statement,
   func-return-list,arg-call-list,func-call-list,
   fcall-ret-statement,decl-list-sp,
   arg-decl-list-sp,arg-decl,decl,
   @pre,@post,specname,boolean-expr,
   scope-decl-list,scope-statement-list,q-list,
   },
}

\begin{figure}[bt]
  \centering
  \begin{minipage}{12cm}
\begin{lstlisting}
/*@\textbf{resolveNonRecursiveFunction($fname$,\Pm)}@*/
 let $\Pm$ be of the form decl-list statement-list
 let type fname(arg-decl-list) { 
       func-decl-list func-body func-return-statement } 
   be /*@\textbf{the non-recursive function declaration}@*/ of $fname$

 // add a return program counter variable to function argument 
 // declarations to hold the label where the function should return 
 append $,~int~\returnpc$ to arg-decl-list
 // add a return variable to the function variable declaration
 append $type~retvar;$ to func-decl-list
 // turn the return statement into an assignment statement 
 append $retvar = expession($func-return-statement$);$ to func-return-list
 // when function done, move program counter back to the caller
 append $pc = \returnpc;$ to func-return-list
 replace func-return-statement with func-return-list
 // iterate over all statements with calls to $\textcolor{darkgreen}{fname}$
 foreach statement $s$ containing an fname(arg-call-list) function call
  // marshal the argument list 
  foreach argument expression $e$ in arg-call-list
   let $a$ be the corresponding variable to $e$ in arg-decl-list
   append $a = e;$ to func-call-list 
  endfor
  // save the label of the statement the function should return to
  // and transfer control to the function 
  append $\returnpc = label($fcall-ret-statement$);$
         $pc = label(first($func-body$));$ to func-call-list
  // declare a unique function call return variable 
  let $i$ be a unique number 
  append $type~\fcallretvari;$ to decl-list
  // create a statement that reads the return value from the function
  let fcall-ret-statement be $\fcallretvari=retvar;$ 
  // place the return point from the function 
  append fcall-ret-statement to func-call-list
  // fix the statement to use the returned value
  $s'$ = substitute fname(arg-call-list) by $fcall\text{-}retvar\text{-}i$ in $s$ 
  append  $s'$ to func-call-list

  replace $s$ with func-call-list
 endfor
\end{lstlisting}
\end{minipage}
\vspace{-1em}
\caption{Resolve non-recursive function calls and declarations}
\label{f:functions}
\vspace{-2em}
\end{figure}

\subsubsection*{Non recursive functions}
\label{s:function}

Algorithm \cci{resolveNonRecursiveFunction} in Figure~\ref{f:functions}
considers the declaration and function calls of a function $fname$. 
It declares two additional variables in the function.
Variable $\returnpc$ is added to \cci{arg-decl-list} and
it holds the label that $pc$ should return to after the function is done.
Variable $retvar$ is added to  \cci{func-decl-list}  and it holds the 
value of the return expression when the function is done.
The \cci{func-return-statement} is then replaced by \cci{func-return-list}
that sets $retvar$ and $pc$ to \cci{expr}(\cci{func-return-statement})
and $\returnpc$, respectively. 

For each statement $s$ containing an $fname$ function call, 
\cci{resolveNonRecursiveFunction}
constructs a list of assignment statements that marshal the arguments
and set the $\returnpc$ to the label of \cci{fcall-ret-statement}.
It then adds an explicit statement that sets $pc$ to the label of the first
statement in $fname$. 
The \cci{fcall-ret-statement} reads the value of $retvar$ into a unique
function call return variable $fcall\textit{-}retvar\textit{-}i$. 
This is necessary to resolve multiple function calls in the same expression. 
Statement $s'$ substitutes the function call by  $fcall\textit{-}retvar\textit{-}i$.
The argument marshaling, \cci{fcall-ret-statement}, and $s'$ form the list 
\cci{func-call-list}.
Finally, statement $s$ is replaced by the list of statements \cci{func-call-list}.

\begin{figure}[bt]
  \centering
  \begin{minipage}{12cm}
\begin{lstlisting}
/*@\textbf{resolveRecursiveFunctionDeclaration($fname$,\Pm)}@*/
 let $\Pm$ be of the form decl-list statement-list
 let type fname(arg-decl-list) { 
       func-decl-list func-body func-return-statement } 
   be /*@\textbf{the recursive function declaration}@*/ of $fname$
 //increase the dimensionality of arguments
 foreach argument declaration arg-decl in arg-decl-list 
  // $\textcolor{darkgreen}{type~var}$ becomes $\textcolor{darkgreen}{type~var[\maxd]}$ and 
  // $\textcolor{darkgreen}{type~array\text{-}var[c]}$ becomes $\textcolor{darkgreen}{type~array\text{-}var[c][\maxd]}$
  append arg-decl$[\maxd],$ to  arg-decl-list-sp
 endfor
 // add a $\textcolor{darkgreen}{\returnpc}$ variable to the argument list
 append $int~\returnpc[\maxd]$ to arg-decl-list-sp
 replace arg-decl-list with arg-decl-list-sp
 // similarly, increase dimensionality for local variables
 foreach declaration decl in func-decl-list 
  append decl$[\maxd];$ to  decl-list-sp
 endfor
 // add return variables
 append $int~retvar[\maxd];$ to decl-list-sp
 replace func-decl-list with decl-list-sp

 // variable  $\textcolor{darkgreen}{\spi}$ plays a stack pointer role for fname
 let $i$ be a unique number 
 append $int~\spi;$ to decl-list

 // starts at negative 1
 // once incremented it is at the first recursive frame
 prepend $\spi= -1;$ to statement-list

 // replace variables with frame variables
 foreach variable $v$ in arg-decl-list and func-decl-list
  replace all occurrences $v$ with $v[\spi]$ in func-body
  replace all occurrences $v$ with $v[\spi]$ in func-return-statement
 endfor
 // similarly for array variables
 foreach array variable $a$ in arg-decl-list and func-decl-list
  replace all occurrences $a[c_1]$ with $a[c_1][\spi]$ in func-body
  replace all occurrences $a[c_1]$ with $a[c_1][\spi]$ in func-return-statement
 endfor
 // fix the return statement to return to the callee
 append $retvar[\spi] = expression($func-return-statement)$;$
        $pc = \returnpc[\spi];$ to func-return-list
 replace func-return-statement with func-return-list
\end{lstlisting}
\end{minipage}
\vspace{-1em}
\caption{Resolve recursive function declarations }
\label{f:recdecl}
\vspace{-2em}
\end{figure}

\subsubsection*{Recursive functions}
\label{s:recfunction}

Algorithm \cci{resolveRecursiveFunctionDeclaration} of
Figure~\ref{f:recdecl} resolves a recursive 
function declaration $fname$ into \psqcore by 
emulating stack frames with depth \maxd. 
The argument and local variables of $fname$ become arrays.
References to them in the body of the function are replaced with
the corresponding array variables indexed by an additional stack pointer
\spi that points to the current recursive depth of the function. 
The rest of the translation follows similarly to Algorithm
\cci{resolveNonRecursiveFunction} of Figure~\ref{f:functions} and 
declares additional variables to hold the return, 
and return program counter values. 

Algorithm \cci{resolveRecursiveFunctionCalls} of Figure~\ref{f:reccall}
%resolves the function calls to a recursive function $fname$. 
works similarly to resolving function calls in 
%\cci{resolveNonRecursiveFunctions} of 
Figure~\ref{f:functions}
with additional attention to the stack pointer \spi. 
Variable \spi is used to 
appropriately marshal the arguments and the return program counter
to the current frame of the recursive function. 
  %if arg-decl is of the form $type arg$
  % append $type~arg[\maxd],$ to arg-decl-list-sp
  %else // arg-decl is of the form type arg[c]
  % append $type~arg[c][\maxd],$ to arg-decl-list-sp
  %endif

  %if decl is of the form type var
  % append $type~var[\maxd];$ to decl-list-sp
  %else // decl is of the form type array-var[c]
  % append $type~array-var[c][\maxd];$ to decl-list-sp
  %endif

%%%%%%%%%%%%%%%%%%%%%%%%%%%%%%%%%%%%%%%%%%%%%%%%%%%%%%%%
\begin{figure}[bt]
  \centering
  \begin{minipage}{12cm}
\begin{lstlisting}
/*@\textbf{resolveRecursiveFunctionCalls(fname, \Pm)}@*/
 let $\Pm$ be of the form decl-list statement-list
 let type fname(arg-decl-list) { 
       func-decl-list func-body func-return-statement } 
   be /*@\textbf{the recursive function declaration}@*/ of $fname$
 let $sp$ be the stack pointer corresponding to $fname$ in decl-list

 foreach statement $s$ containing an fname(arg-call-list) function call 
  // declare a unique function call return variable 
  let $i$ be a unique number 
  append $type~\fcallretvari;$ to decl-list
  // create a statement that reads the return value from the function
  append $\fcallretvari=retvar[sp];$ to fcall-ret-statement
  // marshal the argument list 
  foreach argument expression $e$ in arg-call-list
   let $a$ be the corresponding variable to $e$ in arg-decl-list
   append $a[sp+1] = e;$ to func-call-list 
  endfor
  // save the label of the statement the function should return to
  append $\returnpc[sp+1] = label($fcall-ret-statement$);$ to func-call-list
  // increment the stack pointer 
  append $sp = sp + 1$ to func-call-list
  // transfer control to the function 
  append $pc = label(first($func-body$));$ to func-call-list
  // place the return point from the function 
  append fcall-ret-statement to func-call-list
  // decrement the stack pointer 
  append $sp = sp - 1$ to func-call-list
  // fix the statement to use the returned value
  $s'$ = substitute fname(arg-call-list) by $\fcallretvari$ in $s$ 
  append  $s'$ to func-call-list

  replace $s$ with func-call-list
 endfor
\end{lstlisting}
\end{minipage}
\vspace{-1em}
\caption{Resolve recursive function calls}
\label{f:reccall}
\vspace{-2em}
\end{figure}

%%%%%%%%%%%%%%%%%%%%%%%%%%%%%%%%%%%%%%%%%%%%%%%%%%%%%%%%

\subsubsection*{Specifications and quantifiers}
\label{s:specs}

Algorithm \cci{resolveQuantifier} of Figure~\ref{f:quantifiers}
takes a quantifier expression $q$ and the statement $s$
that contains it. 
It declares a Boolean variable \qi in the scope of the quantifier to 
hold the value of the quantifier. 
It initializes \qi to \true when $q$ is a universal quantifier,
and to \false when $q$ is an existential quantifier. 
It translates the quantifier statement to a loop that iterates over
the range of the quantified variable $v$ and accumulates
the value of the Boolean expression with a conjunction in case 
$q$ was a universal quantifier, and a disjunction in case $q$ 
was an existential quantifier. 
Algorithm \cci{resolveQuantifier} substitutes for $q$ in $s$ by 
\qi to construct $s'$. 
Finally, it replaces $s$ with the constructed loop and $s'$. 
Alternatively, if the Boolean expression in the quantifier
does not include function calls, the quantifier can be unrolled
into a sequence of conjunctions (disjunctions in case of an existential
quantifier) of the Boolean expression for the range of $v$. 
This allows for a smaller execution time.%to execute the quantifier. 
%however, it requires more space as it introduces redundancy of the 
%Boolean expression. 

Algorithm \cci{resolvePrePost} of Figure~\ref{f:specs} 
declares a Boolean variable \cci{pre-specname} 
for each precondition statement $s$ of the form 
\cci{@pre specname \{boolean-expr\}}.
It then replaces the precondition statement 
with an assignment statement that updates \cci{pre-specname} with the 
value of the corresponding Boolean expression \cci{boolean-expr}. 
Algorithm \cci{resolvePrePost} works similarly for the postcondition.
Of special interest is the expression 
$pre\text{-}specname \wedge pc=label(s) \rightarrow post\text{-}specname$
which expresses that the program satisfies the specification.

\begin{lrbox}{2}
\begin{tabular}{p{4.5in}p{0.001in}p{2.9in}}
\begin{lstlisting}
/*@\textbf{resolveQuantifier($q$, $s$, \Pm)}@*/ // statement $\textcolor{darkgreen}{s}$ contains quantifier $\textcolor{darkgreen}{q}$
 let $q$ be of the form $Q$ ($v$ [$e_1$ ... $e_2$]) (boolean-expr) 
 let scope-decl-list and scope-statement-list be the declaration and statement lists of the scope of $s$, respectively
 let $i$ be a unique number // create a unique Boolean variable 
 append $bool~\qi;$ to scope-decl-list
 append $v = e_1;$ to q-list // initialize the quantified variable 
 if $Q$ is a universal quantifier 
  prepend $\qi = \true;$ to scope-statement-list  // initialize $\textcolor{darkgreen}{\qi}$
  // translate the quantifier into a loop
  append $while(v \leq e_2 \land \qi)\{\qi =\qi\land$ boolean-expr$;v=v+1;\}$ to q-list 
 else // $\textcolor{darkgreen}{Q}$ is an existential quantifier 
  prepend $\qi = \false;$ to scope-statement-list 
  append $while(v \leq e_2 \land \neg \qi)\{\qi = \qi \lor$ boolean-expr$;v=v+1;\}$ to q-list 
 endif 
 $s'$ = substitute $q$ by $\qi$ in $s$ // substitute back into the code
 append $s'$ to q-list 
 replace $s$ with q-list 
\end{lstlisting}
& &
\begin{lstlisting}
/*@\textbf{resolvePrePost(\Pm)}@*/
 //declare a Boolean variable for each pre/post condition and replace the statement by an assignment to the declared variable
 foreach precondition statement $s$ of the form (@pre specname{boolean-expr})
  append $bool~pre\text{-}specname;$ to decl-list 
  replace $s$ with $pre\text{-}specname = $boolean-expr$;$
 endfor
 foreach postcondition statement $s$ of the form (@post specname{boolean-expr})
  append $bool~post\text{-}specname;$ to decl-list 
  replace $s$ with $post\text{-}specname = $boolean-expr$;$
 endfor
\end{lstlisting}
\end{tabular}
\end{lrbox}

\begin{figure}[bt]
%  \centering
%\begin{minipage}{12cm}
%\end{minipage}
  \resizebox{1.01\textwidth}{!}{\usebox2} 
\vspace{-1em}
\caption{Resolve quantifiers (left) and pre- and post condition statements (right)}
\label{f:quantifiers}
\label{f:specs}
\end{figure}
%%%%%%%%%%%%%%%%%%%%%%%%%%%%%%%%%%%%%%%%%%%%%%%%%%%%%%%%

%\begin{figure}[bt]
%  \centering
%  \begin{minipage}{12cm}
%\begin{lstlisting}
%/*@\textbf{resolvePrePost(\Pm)}@*/
% //declare a Boolean variable for each and replace the statement
% // by an assignment to the declared variable
% foreach precondition statement $s$ (@pre specname{boolean-expr})
%  append $bool~pre\text{-}specname;$ to decl-list 
%  replace $s$ with $pre\text{-}specname = $boolean-expr$;$
% endfor
% foreach postcondition statement $s$ (@post specname{boolean-expr})
%  append $bool~post\text{-}specname;$ to decl-list 
%  replace $s$ with $post\text{-}specname = $boolean-expr$;$
% endfor
%\end{lstlisting}
%\end{minipage}
%\vspace{-1em}
%\caption{Resolve specifications }
%\label{f:specs}
%\end{figure}
%

\subsection{\oneloop to AIG circuits }
\label{s:onelooptoaig}
\lstset{
   keywords=[1]{for,foreach,with,endfor,in,if,elseif,endif,to,else,let,be,by,from,
   substitute,replace,return},
   keywords=[2]{vargates,type,append,array-access,expr,
     next-list,decl-list,init-list,statement-list,array-expr,a-expr-resolve,a-resolve,
   },
}

\begin{figure}
  \begin{tabular}{p{6.5cm}p{9cm}}
\begin{lstlisting}
/*@\textbf{traverse}@*/($exp$)
 // base case of recursion
 if ($exp$ is a variable) 
  return vargates($exp$)
 elseif ($exp$ is constant array access)
  return vargates($exp$)
 elseif ($exp$ is array access) 
  return /*@\textbf{resolveArrayAccess}@*/($exp$)
 endif
 // traverse operands and store results in $\textcolor{darkgreen}{wirevec}$
 for $i$ from $1$ to $exp.operands.size()$ 
  $wirevec[i]$ = /*@\textbf{traverse}@*/($exp.operands[i]$) 
 endfor
 //lookup AIG circuit for operation in library
 return library($exp.operation$, $wirevec$)
\end{lstlisting}
&
\begin{lstlisting}
/*@\textbf{variables}@*/(decl-list)
 foreach variable $v$ in decl-list
  if $v$ is an array variable
   //handle array variables
   for $j$ from 0 to size($v$) - 1
    vargates($v[j]$) = 
      instantiate-registers($v[j]$,type($v[j]$))
   endfor
  elseif ($v$ is a primary input)
   // handle nondeterministic variables
   vargates($v$) = 
     instantitate-primary-inputs($v$,type($v$))
  else 
   vargates($v$) = 
     instantiate-registers($v$,type($v$))
  endif
 endfor
\end{lstlisting}
\end{tabular}
\vspace{-2em}
\caption{Expression traversal and register instantiations for variables}
\label{f:traverse}
\end{figure}

Algorithm \cci{variables} of Figure~\ref{f:traverse}
instantiates vectors of AIG registers and primary inputs to translate the 
corresponding \oneloop program variables and stores the translation in 
a lookup function \cci{vargates}. 
The width of a bit vector can be selected by the user, 
or set to match the default width of the declared type. 
Typically the default values for the bit width are 
$32$ bits for an integer.
Arrays are represented by a fixed number of array elements
and each element is then treated as a regular variable. 
The number of array elements is either specified in the program, 
bounded by the user, or fixed to a constant by \psqtool. 

\subsubsection{Assignment statements}
\psqtool considers each assignment statement $s$ in \cci{init-list}
and \cci{next-list} and traverses the right hand side expression of
$s$ with the recursive \cci{traverse} Algorithm of Figure~\ref{f:traverse}. 
If expression $exp$ refers to a variable (base case), 
or an array access operator with a constant index, 
then \cci{traverse} returns \cci{vargates($exp$)}. 
If $exp$ was an array access $a[ie]$ where $a$ is the array
variable and $ie$ is an index expression, then \cci{resolveArrayAccess}
of Figure~\ref{f:array}
translates $exp$ into a nested ternary expression where the conditions
depend on the value of $ie$ and the values are array access terms 
with constant indices between $0$ and $size(a)-1$. 
Algorithm \cci{resolveArrayAccess} calls \cci{traverse} again on the
generated expression to produce the corresponding AIG. 

If the expression is a logical, conditional, or arithmetic expression, then
the \cci{library} routine looks it up in a complete table of AIG circuits
with the adequate bit width. 
For example, if the expression is a ternary conditional statement of the
form $b~?~e_1:e_2$, then \cci{library} instantiates a multiplexer, 
connects its two data fanins to the nodes corresponding to $e_1$ and $e_2$, 
connects its control fanins to the nodes corresponding to $b$,
and returns its fanouts. 
Alternatively, users have the choice to abstract 
multiplication, division, and remainder by non-constant factors with non-deterministic 
variables (uninterpreted functions). 

\subsubsection{Connections and array access target terms}
For assignment statements with target terms that are regular variables, 
or array access with constant index expressions, 
\psqtool connects the nodes corresponding to the right hand side 
expressions in \cci{init-list} and 
\cci{next-list} to the initial and next state value 
fanins of the corresponding register gates, 
respectively. 

An assignment statement $s$ of the form 
$a[ie] = expr;$, where $a$ is an array variable,
$ie$ is a non constant index expression, and $expr$ is an expression,
requires preprocessing before being translated to AIG.
Algorithm \cci{variables} instantiates $size(a)-1$ register vectors 
corresponding to the elements of array variable $a$, 
each requiring initial and next state value functions. 
Algorithm \cci{resolveArrayTargetTerms} of Figure~\ref{f:array}
takes as input an assignment statement $s$
and replaces it with $size(a)-1$ assignment statements 
of the form $a[j]=(j==ie)?expr:a[j];$
where $j$ ranges from $0$ to $size(a)-1$ such that each 
array element $a[j]$ evaluates to $expr$ if $ie$ evaluates to $j$,
otherwise, $a[j]$ keeps its value. 

\begin{figure}[bt]
  \centering
\begin{tabular}{p{6cm}p{9cm} }
\begin{lstlisting}
/*@\textbf{resolveArrayAccess($exp$)}@*/
 let $a$ be base($exp$)
 let $ie$ be index($exp$) 
 let $N$ be size(a) - 1 

 for $j$ from $0$ to $N-1$
  append $(ie == j)~?~a[j]~:$ to a-expr-resolve
 endfor 
 append $a[N]$ to a-expr-resolve
 return /*@\textbf{traverse}@*/($exp$)
\end{lstlisting}
&
\begin{lstlisting}
/*@\textbf{resolveArrayTargetTerms(\Pm)}@*/
 foreach statement $s$ where target(s) is array-access
  let $a$ be base(array-access)
  let $ie$ be index(array-access) 
  let $N$ be size (a) - 1 

  for $j$ from $0$ to $N$
   append $a[j] = (j == ie)~?~$expr$~:~a[j];$ to a-resolve
  endfor 
  replace $s$ with a-resolve
 endfor
\end{lstlisting}
\end{tabular}
\vspace{-2em}
\caption{Resolve array access expressions and resolve array target terms}
\label{f:array}
\end{figure}

%%%%%%%%%%%%%%%%%%%%%%%%%%%%%%%%%%%%%%%%%%%%%%%%%%%%%%
%%%%%%%%%%%%%%%%%%%%%%%%%%%%%%%%%%%%%%%%%%%%%%%%%%%%%%

%\section{Termination guarantee bound}
%\label{s:termination}
%\input{termination}

%%%%%%%%%%%%%%%%%%%%%%%%%%%%%%%%%%%%%%%%%%%%%%%%%%%%%%

\section{Implementation}
\label{s:implementation}
\begin{figure}[tb]
  \centering
  \resizebox{.9\textwidth}{.25\textheight}{
    \includegraphics{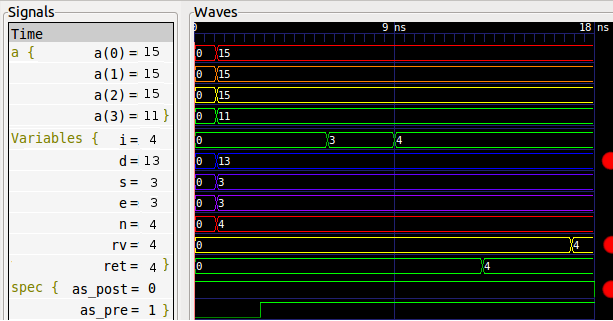}
  } 
\caption{Array search visual debugging with GTKWave~\cite{gtkwave}.}
\label{fig:vdebug}
\end{figure}

We fully implemented \psqtool using C++ and ANTLR~\cite{parr1995antlr} and 
integrated our implementation
with the ABC synthesis and verification framework~\cite{brayton2010abc}
and the GTKWave visualization tool~\cite{gtkwave}.
The frontend supports the \psqlanguage syntax under C, C++, and Java. 
Figure~\ref{fig:vdebug} shows a trace that highlights the defect of the 
array search program of Figure~\ref{fig:arraysearch} where $d$ is not
in $a$ and $rv$ is not $-1$. 
\psqtool provides a fully automated verification path of execution where it executes 
a well selected sequence of synthesis reduction and verification techniques that
work well for software verification tasks. 
\psqtool also supports an interactive shell with parser, synthesis, verification,
and debugging commands where the user can save and use intermediary results, 
try different synthesis and verification strategies, and inspect counterexamples. 
Table~\ref{tb:command:reference} provides a short list of the 
\psqtool commands. 
The commands allow the users to specify (1) the program, 
(2) bounds on variable data width, array size, recursion depth, 
and (3) optional automatically embedded checks on array out of bound access
and arithmetic overflow. 
\psqtool is available online as an open source tool with tutorial and 
documentation~\footnote{\label{fn:online}\url{http://research-fadi.aub.edu.lb/dkwk/doku.php?id=sa}}.

\begin{table}[tb]
\caption{Summary of \psqtool{} commands}
\centering
\resizebox{.96\textwidth}{!}{
\begin{tabular}{|p{0.16\textwidth}|p{1.1\textwidth}|}
\hline
\multicolumn{1}{|c}{{\bf Command}} & \multicolumn{1}{|c|}{{\bf Description}} \\ \hline \hline
\cci{read} & Parses a program and generates a parse
graph. \\ \hline
%It returns error messages in case parsing errors were found \\ \hline
%\cci{print\_stats} & Prints some statistics about the parsed code graph \\ \hline
%\cci{print\_conf} & Prints the default synthesis configuration entries. \\ \hline
\cci{print-time} & Prints the runtime time of the last command and the total run time. 
\\ \hline
\cci{prove-scheme1} & Performs a predefined scheme of sequential synthesis and
proof strategies. \\ \hline
\cci{prove-scheme2} & Performs a predefined scheme of sequential synthesis and
proof strategies. \\ \hline
\cci{abc cmd} & Performs the ABC techniques specified in \cci{cmd}.\\ \hline
\cci{debug} & Generates a program counterexample from the AIG counterexample
and allows the user to inspect the values in a textual format. \\ \hline
\cci{vdebug} & A visual version of \cci{debug} that allows the user to view
and interact with the traces using GTKWave. \\ \hline 
\cci{start-sim} & Starts the simulator helper tool to help the user analyze the code 
and perform what-if analysis on variable values. 
\\ \hline
\cci{sim-variable} & Runs the simulator helper tool to generate and display 
values for specific variables. \\ \hline
\cci{start-prove} & Configures the prove engine with parameters such as 
variable bit width, array size bounds, and recursion depth bounds, as well as
overflow check, out of bound access check, and quantifier unrolling. 
\\ \hline
%
%
%\cci{stop\_prove} & Stops the current proof session and deletes the generated
%AIG \\ \hline
%
%\cci{stop\_sim} & Stops the current simulation session and deletes the generated 
%AIG \\ \hline
%\cci{abc cmd} & Call ABC with the command \cci{cmd} \\ \hline
% \cci{help} & Display a list of available commands \\ \hline
% \cci{print\_conf} & Print the current configuration of \psqtool{} \\ \hline
%\cci{prove\_status} & Print the status of the current proving session \\ \hline
%\cci{sim\_status} & Print the status of the current simulation session \\ \hline
%\cci{start\_prove} & Start a proving session \\ \hline
% \cci{start\_sim} & Start a simulation session \\ \hline
%\cci{sim\_variable} & Print variable values for a range of simulation frames \\ \hline
%\cci{stop\_prove} & Stop the current proving session \\ \hline
%\cci{stop\_sim} & Stop the current simulation session \\ \hline
%\cci{debug} & Debug the counter example generated from the current proving session \\ \hline
%\cci{quit,exit} & Exit the \psqtool{} environment \\ \hline
\end{tabular}
}
\label{tb:command:reference}
\end{table}

%%%%%%%%%%%%%%%%%%%%%%%%%%%%%%%%%%%%%%%%%%%%%%%%%%%%%%
%%%%%%%%%%%%%%%%%%%%%%%%%%%%%%%%%%%%%%%%%%%%%%%%%%%%%%
%\todo{check on this and whether anything should remain from it}
%\section{Sequential circuit encoding}
%\label{s:encode}
%\input{scencode}

%%%%%%%%%%%%%%%%%%%%%%%%%%%%%%%%%%%%%%%%%%%%%%%%%%%%%%
%%%%%%%%%%%%%%%%%%%%%%%%%%%%%%%%%%%%%%%%%%%%%%%%%%%%%%

\section{Results}
\label{s:results}
We evaluated \psqtool by verifying software artifacts 
including library algorithms and data structures with complete sophisticated specifications 
from the Calculus of Computation book~\cite{MannaBradleyCalculus2007},
%a sophisticated double array trie data structure (DATRIE) with partial 
%specifications taken from assertions and formal descriptions~\cite{datrie89}, 
an array based implementation of the red black balanced binary search 
tree (RBBST)%~\cite{RBBST2010} 
with specifications 
formalized from~\cite{CormenETAL90IntroductionAlgorithms}, and 
a memory allocation model (MMAN) with complete specifications 
provided from the Frama-C ANSI C Specification Language (ACSL)~\cite{ACSL15}.
%and benchmark applications (GZIP,TOKENS,TCAS) 
%from the SIR repository ~\cite{Do2005}
%with partial specifications provided from~\cite{PBCOV2014}. 
Table~\ref{t:standard} reports the results 
%software artifacts with formal specifications 
with several bounds and compares against CBMC\cite{clarke2004tool}. 
The results show that \psqtool scales to bounds larger than CBMC
by at least one order of magnitude. 

We also evaluated \psqtool with benchmark verification tasks
from the 2014 software verification competition~\cite{SVComp2014}
each with memory safety, termination, and error label reachability checks. 
Some of the tasks are \true where a proof is expected, and some
are \false where a counterexample is expected. 
We ran our experiments on a machine with similar settings to the server machine reported in~\cite{SVComp2014}
(3.4 GHz 64-bit Quad Core CPU, a 64 bit GNU/Linux
operating system, and 32 GB of RAM).
We compare our results to the leading tools in the competition. 
Table~\ref{t:svcomp1} reports the results 
where some of the leading tools in the competition 
(CBMC~\cite{clarke2004tool}, ESBMC~\cite{ESBMC2014TACAS}, BLAST~\cite{BLAST-2007Beye}, 
and CPAChecker~\cite{CPACheckerFramework2012}) 
successfully completed the verification task and compares against them in terms of running time. 
The results show that \psqtool succeeded to verify all the tasks
and ranked either first or second in terms of running time in
comparison with the leading tools. 
Table~\ref{t:svcomp2} reports the results of verifying 
benchmarks from ~\cite{SVComp2014} where the leading tools 
reported a 
timeout and failed to complete the verification task. 
According to the wrapper scripts provided from~\cite{SVComp2014},
CBMC and ESBMC report a \true result by default when 
they reach their timeout while the SAT or SMT solver are 
running, they report an $\mathit{unknown}$ result when the 
timeout passes before they generate the first CNF or SMT
instance to pass to the underlying solver. 
This is {\em probably reasonable}  as CBMC and ESBMC 
consider the 
inability of the solvers to find counterexamples within the 
timeout period as an indication of the rarity or absence 
of such results. 
Thus we consider a \true result with a timeout time from
CBMC and ESBMC as a timeout result for comparison purposes. 
The results show that \psqtool succeeded to complete 
verification tasks that were not completed by 
the leading tools.

In the experiments, we used an automated script that ran reduction algorithms 
first with a timeout of 10 minutes with Table~\ref{t:svcomp1} and three minutes
for Tables~\ref{t:svcomp1} and ~\ref{t:svcomp2}, 
followed by three instances of verification using the \cci{pdr}, \cci{dprove},
and \cci{ind} ABC commands with proper timeout, number of frames, and BDD 
size configurations.

\begin{table}
\centering 
\caption{ Results of \psqtool and CBMC to verify array based algorithms and data structures with increasing array
  sizes. TO, MO, ERR, MISTAKE stand for timeout, memory out, runtime error, and verification mistake, respectively.
}
\label{t:standard}
\resizebox{.95\textwidth}{!}{
\begin{tabular}{|l|c|| c|c|c|| c|c |p{0.001cm}| c|c|| c|} \cline{2-7} \cline{9-11}
 \multicolumn{1}{c|}{}
 &  Array & \multicolumn{3}{|c||}{AIG size before/after reductions} & \multicolumn{2}{|c|}{\psqtool} &  & \multicolumn{2}{|c||}{CNF size  }  & CBMC \\ \cline{3-7} \cline{9-11}

\multicolumn{1}{c|}{}
 & size & registers& and & level & verif. & total (s) &  & vars & clauses & time (s) \\
   \cline{2-7} \cline{9-11} 
   \cline{1-7} \cline{9-11} 
array & 3 &   86 / 41  &   719 / 313  &   24 / 15  & 0.33 & 4.36 &  & 2,416 & 6,784 & 0.016 \\ \cline{2-7} \cline{9-11}
search & 7 &   118 / 68  &   1,064 / 568  &   27 / 19  & 3.89 & 12.4 &  & 4,612 & 15,008 & 722.4 \\ \cline{2-7} \cline{9-11}
 & 15 &   174 / 119  &   1,781 / 1,116  &   30 / 21  & 2.41 & 16.87 &  & 9,112 & 34,496 &  TO \\ \cline{2-7} \cline{9-11}
 & 31 &   286 / 226  &   3,362 / 2,346  &   45 / 24  & 1.43 & 33.67 &  & 18,332 & 84,928 &  TO \\ \cline{2-7} \cline{9-11}
 & 63 &   526 / 461  &   6,895 / 5,100  &   78 / 26  & 4.57 & 99.64 &  & 37,216 & 230,208 &  TO \\ \cline{2-7} \cline{9-11}
 & 127 &   1,054 / 984  &   14,780 / 11,315  &   143 / 28  & 21.32 & 397 &  & 75,876 & 695,616 &  TO \\ \cline{2-7} \cline{9-11}
 & 255 &   4,798 / 4,718  &   70,742 / 55,364  &   529 / 33  & 682.1 & 8,022.1 &  & TO & TO & TO \\ 
   \hline \hline
binary 
 & 3 &   94 / 56  &   879 / 555  &   30 / 19  & 0.11 & 1.04 &  & 6,503 & 24,533 & 0.085 \\ \cline{2-7} \cline{9-11}
search 
 & 7 &   151 / 83  &   1,832 / 850  &   42 / 17  & 0.54 & 1.47 &  & 16,172 & 68,130 & 1.91 \\ \cline{2-7} \cline{9-11}
 & 15 &   268 / 143  &   5,185 / 1,943  &   62 / 20  & 25.42 & 27.69 &  & 42,461 & 197,223 & 38.493 \\ \cline{2-7} \cline{9-11}
 & 31 & 491/262 & 18,355/4,903 & 68/28 & 112.84 & 115.13 &  & 100,379 & 412,410 & 4,955.223 \\ \cline{2-7} \cline{9-11}
 & 63 & 874/454 & 64,457/8,608 & 102/28 & 1,132 & 1,152.22 &  & 2,660,677 & 13,689,632 &  TO \\ \cline{2-7} \cline{9-11}
 & 127 & 1,911/976 & 277,662/19,907 & 170/32 & TO & TO &  & TO & TO & TO \\ 
   \hline \hline
bubble 
 & 3 &   114 / 44  &   1,198 / 393  &   29 / 16  & 0.29 & 5.79 &  & 15,534 & 43,332 & 0.174 \\ \cline{2-7} \cline{9-11}
sort 
 & 7 &   169 / 68  &   2,218 / 885  &   35 / 20  & 17.1 & 31.09 &  & 63,785 & 197,603 & 7.298 \\ \cline{2-7} \cline{9-11}
 & 15 &   276 / 117  &   5,607 / 2,106  &   47 / 22  & 1,390.25 & 1,426.98 &  & 341,552 & 1,082,480 & 455.5 \\ \cline{2-7} \cline{9-11}
 & 31 & 603/360 & 40,448/25,546 & 75/36 & 2,668.22 & 2,669.82 &  & 2,144,801 & 6,809,622 & TO \\ \cline{2-7} \cline{9-11}
 & 63 & 1,293/814 & 174,535/110,123 & 140/40 & 4,323.51 & 4,384.08 &  & TO & TO & TO \\ \cline{2-7} \cline{9-11}
 & 127 & 2,847/1,844 & 770,398/485,523 & 269/44 & TO & TO &  & T0 & TO & TO \\ 
   \hline \hline
selection 
 & 3 &   86 / 45  &   1,021 /679  &   26 / 19  & 20.62 & 20.69 &  & 17,941 & 55,151 & 0.37 \\ \cline{2-7} \cline{9-11}
sort 
 & 7 &   150 / 87  &   2,726 / 1,971  &   34 / 24  & 1,105.2 & 1,105.74 &  & 315,505 & 1,129,784 & TO \\ \cline{2-7} \cline{9-11}
 & 15 &   280 / 125  &   5,676 / 2,236  &   47 / 22  & 2,280 & 2,280.18 &  & 959,841 & 3,849,448 & TO \\ \cline{2-7} \cline{9-11}
 & 31 & 590/383 & 39,543/26,547 & 75/27 & TO & TO &  & 3,269,941 & 14,305,256 & TO \\ \cline{2-7} \cline{9-11}
 & 63 & 1,278/841 & 172,385/112,601 & 140/31 & TO & TO &  & TO & TO & TO \\
   \hline \hline
array 
 & 3 &   110 / 57  &   1,896 / 689  &   33 / 19  & 0.87 & 2.79 &  & 36,636 & 121,478 & 2.43 \\ \cline{2-7} \cline{9-11}
partition
 & 7 &   147 / 87  &   2,509 / 1,174  &   34 / 24  & 93.47 & 97.56 &  & 622,799 & 2,663,342 & TO \\ \cline{2-7} \cline{9-11}
 & 15 &   206/ 138  &   3,819 / 2,651  &   38 / 35 & 2,127 & 2,135.7 &  & 2,061,314 & 9,390,063 & TO \\ \cline{2-7} \cline{9-11}
 & 31 & 323/248 & 6,779/5,557 & 42/39 & TO & TO &  & 7,315,089 & 34,755,647 & TO \\ \cline{2-7} \cline{9-11}
 & 63 & 568/486 & 13,451/12,178 & 46/43 & TO & TO &  & TO & TO & TO \\ \cline{2-7} \cline{9-11}
   \hline \hline
linked 
 & 3 &   237 / 98  &   4,310 / 871  &   38 / 19  & 109.63 & 118.89 &  & \color{red}{28,347} & \color{red}{28,347} & \color{red}{0.194 (mistake)} \\ \cline{2-7} \cline{9-11}
list
 & 7 &   344 / 179  &   6,117 / 1,693  &   41 / 26  & 1,800 & 1,811 &  & \color{red}{83,079} & \color{red}{83,079} & \color{red}{0.527 (mistake)} \\ \cline{2-7} \cline{9-11}
insert 
 & 15 & 503/345 & 10,521/3,763 & 47/26 & TO & TO &  & \color{red}{281,759} & \color{red}{281,759} & \color{red}{2.027 (mistake)} \\ \cline{2-7} \cline{9-11}
 & 31 & 839/634 & 19,624/10,060 & 53/36 & TO & TO &  & \color{red}{1,039,655} & \color{red}{1,039,655} & \color{red}{13.575 (mistake)} \\ \cline{2-7} \cline{9-11}
 & 63 & 1,559/1,482 & 40,375/16,396 & 59/32 & error & error &  & \color{red}{3,946,601} & \color{red}{3,946,601} & \color{red}{152.92 (mistake)} \\
   \hline \hline
linked 
 & 3 &   197 / 84  &   2,906 / 722  &   33 / 21  & 47.72 & 71.383&  & \color{red}{22,898} & \color{red}{22,898} & \color{red}{0.163 (mistake)} \\ \cline{2-7} \cline{9-11}
list 
 & 7 &   293 / 157  &   4,454 / 1,387  &   39 / 25  & 1,800.15 & 1,830.21 &  & \color{red}{59,219} & \color{red}{59,219} & \color{red}{0.457 (mistake)} \\ \cline{2-7} \cline{9-11}
remove 
 & 15 & 448/397 & 7,847/3,094 & 45/20 & TO & TO &  & \color{red}{175,176} & \color{red}{175,176} & \color{red}{1.935 (mistake)} \\ \cline{2-7} \cline{9-11}
 & 31 & 778/821 & 15,763/6,364 & 51/23 & TO & TO &  & \color{red}{579,957} & \color{red}{579,957} & \color{red}{11.163 (mistake)} \\ \cline{2-7} \cline{9-11}
 & 63 & 1,492/1,584 & 233,791/14,346 & 57/25 & TO & TO &  & \color{red}{2,038,358} & \color{red}{2,038,358} & \color{red}{85.94 (mistake)} \\
   \hline \hline
RBBBST 
 & 3 & 779/407 & 14,284/6,451 & 122/41 & 313.11 & 314.22 &  & \color{red}{2,765,445} & \color{red}{2,765,445} & \color{red}{7.169 (mistake)} \\ \cline{2-7} \cline{9-11}
insert 
 & 7 & 1,555/1,077 & 40,794/29,281 & 128/54 & 3,600.48 & 4,087.1 &  & \color{red}{5,602,125} & \color{red}{5,602,125} & \color{red}{15.456 (mistake)} \\ \cline{2-7} \cline{9-11}
 & 15 & 1,834/1,221 & 50,540/28,928 & 134/43 & TO & TO &  & 10,487,436 & 38,445,954 & TO \\ \cline{2-7} \cline{9-11}
 & 31 & 2,810/2,124 & 100,841/66,558 & 152/49 & TO & TO &  & 21,680,107 & 21,680,107 & TO \\ \cline{2-7} \cline{9-11}
 & 63 & 4,789/4,034 & 213,576/147,577 & 186/56 & TO & TO &  & MEMOUT & MEMOUT & MEMOUT \\
   \hline \hline
RBBBST
 & 3 & 1,487/923 & 34,506/15,309 & 251/44 & 124.63 & 313.51 &  & \color{red}{4,403,782} & \color{red}{4,403,782} & \color{red}{8.57 (mistake)} \\ \cline{2-7} \cline{9-11}
remove 
 & 7 & 1,559/993 & 43,608/18,963 & 253/41 & 3,600.62 & 4,154.93 &  & \color{red}{8,587,721} & \color{red}{8,587,721} & \color{red}{17.972 (mistake)} \\ \cline{2-7} \cline{9-11}
 & 15 & 2,072/1,404 & 69,646/48,341 & 254/56 & TO & TO &  & \color{red}{17,476,396} & \color{red}{17,476,396} & \color{red}{44.67 (mistake)} \\ \cline{2-7} \cline{9-11}
 & 31 & 2,969/2,308 & 139,461/101,039 & 259/53 & TO & TO &  & \color{red}{37,383,755} & \color{red}{37,383,755} & \color{red}{327.682 (mistake)} \\ \cline{2-7} \cline{9-11}
 & 63 & 4,610/3,901 & 291,311/233,674 & 262/59 & TO & TO &  & MEMOUT & MEMOUT & MEMOUT \\ 
   \hline \hline
memory 
% &  &  &  &  & 7,959.46 & 80,145.2 &  &  &  &  \\ \cline{2-7} \cline{9-11}
 & 16 & 3,033/962 & 41,137/13,461 & 366/31 & 460.09 & 608.61 &  & TO & TO & TO \\ \cline{2-7} \cline{9-11}
manager
 & 64 & 8,799/2,873 & 99,510/52,543 & 1,072/36 & 1,759.52 & 2,238.18 &  & TO & TO & TO \\
   \hline \hline
   \multicolumn{2}{|c||}{average reductions} & 64.5\% & 51.2\% & 44.7\% & \multicolumn{6}{c}{} \\ \cline{1-5} 
\end{tabular}
}
\end{table}

\begin{table}
\centering 
\caption{ Runtime results in seconds for verification tasks completed successfully by \psqtool compared to the leading software verification competition tools. TO and ERR stand for timeout and error, respectively. }
\label{t:svcomp1}
\resizebox{.8\textwidth}{!}{
  \begin{tabular}{|l|c||c||c|c|c|c|} \cline{2-7}
  \multicolumn{1}{c|}{} 
 &Category &\psqtool&BLAST&CBMC&CPAChecker&ESBMC \\ \hline \hline 
Problem01\_00\_true&eca&7.5&TO&TO-T&4.5&4.6 \\ \hline
Problem01\_10\_true&eca&8.59&730&TO-T&4.5&4.2 \\ \hline
Problem01\_30\_true&eca&9.06&TO&TO-T&4.6&4 \\ \hline
Problem01\_40\_true&eca&9.41&TO&TO-T&4.5&5.9 \\ \hline
Problem02\_00\_true&eca&10.68&460&TO-T&4.3&6.3 \\ \hline
Problem02\_10\_true&eca&9.71&580&TO-T&4.3&4.8 \\ \hline
Problem02\_20\_true&eca&10.14&ERR&TO-T&4.3&4.8 \\ \hline
Problem01\_20\_false&eca&9.35&TO&150&5.3&33 \\ \hline
Problem01\_50\_false&eca&9.87&TO&22&4.8&33 \\ \hline
Problem01\_60\_false&eca&5.38&780&0.56&2.8&30 \\ \hline \hline
s3\_clnt\_1\_true&ssh-simplified&57.3&140&88&3.5&3.4 \\ \hline
s3\_clnt\_1\_false&ssh-simplified&47.15&23&2.1&2.8&13 \\ \hline
s3\_clnt\_2\_true&ssh-simplified&47.1&TO&90&3.6&4.2 \\ \hline
s3\_clnt\_2\_false&ssh-simplified&44.26&23&2.1&2.8&14 \\ \hline
s3\_clnt\_3\_true&ssh-simplified&77.64&350&110&3.7&3.9 \\ \hline
s3\_clnt\_3\_false&ssh-simplified&27.56&23&2.5&2.8&17 \\ \hline
s3\_clnt\_4\_true&ssh-simplified&79.45&TO&90&3.5&3.6 \\ \hline
s3\_clnt\_4\_false&ssh-simplified&76.95&23&2.1&2.8&14 \\ \hline
s3\_srvr\_1a\_true&ssh-simplified&20.11&1.2&3.4&1.8&0.69 \\ \hline
s3\_srvr\_1b\_true&ssh-simplified&1.68&1.2&1.3&1.6&0.48 \\ \hline
s3\_srvr\_1\_false&ssh-simplified&35.47&4.2&3&2.4&22 \\ \hline
s3\_srvr\_2\_true&ssh-simplified&567.51&410&200&4.4&3.9 \\ \hline
s3\_srvr\_2\_false&ssh-simplified&37.05&4.4&2.8&2.4&18 \\ \hline
s3\_srvr\_4\_true&ssh-simplified&799&440&210&14&3.8 \\ \hline
s3\_srvr\_6\_true&ssh-simplified&694&TO&230&24&4.3 \\ \hline
s3\_srvr\_6\_false&ssh-simplified&14.24&0.15&0.21&19&21 \\ \hline
s3\_srvr\_8\_true&ssh-simplified&340&220&220&4.7&4.1 \\ \hline \hline
test\_locks\_5\_true&ControlFlowInt&0.15&23&1.6&2.1&0.43 \\ \hline
test\_locks\_6\_true&ControlFlowInt&0.23&44&2.4&2.7&0.42 \\ \hline
test\_locks\_7\_true&ControlFlowInt&0.31&140&2.8&5&0.45 \\ \hline
test\_locks\_8\_true&ControlFlowInt&0.37&280&3.6&21&0.48 \\ \hline
test\_locks\_9\_true&ControlFlowInt&0.55&610&4.5&53&0.52 \\ \hline
test\_locks\_10\_true&ControlFlowInt&0.84&TO&5.6&53&0.54 \\ \hline
test\_locks\_11\_true&ControlFlowInt&1.24&TO&7&53&0.6 \\ \hline
test\_locks\_12\_true&ControlFlowInt&0.99&TO&8.7&53&0.62 \\ \hline
test\_locks\_13\_true&ControlFlowInt&1.14&TO&11&52&0.66 \\ \hline
test\_locks\_14\_true&ControlFlowInt&1.56&TO&14&52&0.68 \\ \hline
test\_locks\_15\_true&ControlFlowInt&1.83&TO&15&53&0.77 \\ \hline
test\_locks\_14\_false&ControlFlowInt&0.46&0.14&0.22&2&1.4 \\ \hline
test\_locks\_15\_false&ControlFlowInt&0.51&0.13&0.21&2&1.5 \\ \hline 
\end{tabular}
}
\end{table}

%&yellow&exception&&&&
%&red&timeout&&&&

\begin{table}
\centering 
\caption{ Results for verification tasks in seconds where 
  \psqtool terminated successfuly while the leading tools 
  either timed out or produced erroneous output. 
TO: timeout, TO-T: report true on timeout, TO-F report false on timeout, MISTAKE: report erroneous results, ERR: reported a runtime error, and UNKNOWN: report inconclusive result. }
\label{t:svcomp2}
\resizebox{.97\textwidth}{!}{
  \begin{tabular}{|l|c||c||c|c|c|c|} \cline{2-7}
  \multicolumn{1}{c|}{} 
  &Category &\psqtool&BLAST&CBMC&CPAChecker&ESBMC\\ \hline \hline
Problem07\_00\_true&ControlFlowInt/eca&620&TO&TO-T&TO&UNKNOWN-TO\\ \hline
Problem09\_00\_true&ControlFlowInt/eca&433&TO&UNKNOWN-TO&TO&UNKNOWN-190\\ \hline
Ackermann01\_true&Recursive&723&N/A&TO-T&ERR-1.9&MISTAKE-TO\\ \hline
Ackermann02\_false&Recursive&22&N/A&0.73&ERR-1.9&TO-F\\ \hline
Addition03\_false&Recursive&11.2&N/A&MISTAKE-TO&ERR-2&TO\_F\\ \hline
EvenOdd01\_true&Recursive&0.68&N/A&0.96&ERR-1.9&MISTAKE-0.4\\ \hline
EvenOdd03\_false&Recursive&0.45&N/A&0.2&ERR-1.9&0.4\\ \hline
insertion\_sort\_true&loops&154&MISS-6.6&TO-T&TO&UNKNOWN-TO\\ \hline
invert\_string\_false&loops&0.22&1.8&0.47&TO&0.6\\ \hline
jain\_5\_true&bitvector&1.44&N/A&0.53&TO&0.25\\ \hline
linear\_search\_false&loops&0.39&0.13&0.62&ERR-2.6&MISTAKE-0.3\\ \hline
McCarthy91\_false&Recursive&1.2&N/A&0.19&ERR-1.9&TO-F\\ \hline
McCarthy91\_true&Recursive&567&N/A&TO-T&ERR-1.9&MISTAKE-TO\\ \hline
MultCommutative\_true&Recursive&780&N/A&TO-T&ERR-1.9&MISTAKE-TO\\ \hline\hline

s3\_srvr\_1\_alt\_true.BV&bitvector&TO&N/A&TO-T&TO&TO-T\\ \hline
s3\_srvr\_1\_true&ssh-simplified&TO&UNKNOWN-0.1&160&4.6&3.9\\ \hline
s3\_srvr\_3\_true&ssh-simplified&TO&430&210&13&4\\ \hline
s3\_srvr\_7\_true&ssh-simplified&TO&TO&210&24&4.3\\ \hline
Problem08\_00\_true&ControlFlowInt/eca&TO&TO&UNKNOWN-TO&TO&UNKNOWN-TO\\ \hline
\end{tabular}
}
\end{table}

\subsection{Algorithms and data structures}

Table\ref{t:standard} shows that \psqtool was able to verify 
programs with sophisticated specifications faster than CBMC
and for array sizes that CBMC could not verify. 
The table shows the number of memory elements (registers), gates, 
and logic levels before and after the ABC reductions. 
It also shows the total time taken and the time taken to verify the programs. 
The CBMC columns show the size of the generated CNF formulae in terms of the 
number of CNF variables and clauses. 
We developed the programs and the specifications and called
\psqtool and CBMC to verify their correctness incrementally. 
Both \psqtool and CBMC returned counterexamples that we used to correct
the programs and the specifications. 
For programs with array indirection where array elements 
are used as indexes of other arrays such as \cci{next}, 
and \cci{left}, \cci{right}, and \cci{parent} in the linked 
list and red black balanced binary search tree (RBBBST), 
\psqtool and CBMC both succeeded in returning counterexamples for index 
out of bound violations.
However, once we corrected those issues, 
CBMC mistakenly reported that the programs are correct while 
\psqtool correctly reported counterexamples related to the cardinality of
the data structures after insertion or removal of elements. 
The rows highlighted in red report the time taken by CBMC to verify the 
defected code and return a wrong result versus the time taken by \psqtool 
to verify the correct code. 
Note that CBMC and other tools reportedly returned several wrong results for 
benchmarks in the 2014 software verification competition~\cite{SVComp2014},
\psqtool performed better than CBMC as follows.
\begin{itemize}
  \item Most of the times where both \psqtool and CBMC succeeded to 
    verify the programs, \psqtool took less time to complete the verification
    task.
  \item \psqtool succeeded to verify the programs for bounds larger than 
    what CBMC could verify within the three hours seconds timeout. 
  \item CBMC failed to generate the CNF formula for some programs
    with relatively larger array sizes while \psqtool succeeded to 
    verify some of them and generated AIG circuits but timed out
    while verifying the rest. 
    This is significant since 
    model checking {\em typically} returns counterexamples fast when they exist, and
    takes long time to report a proof for unsatisfiable (correct) claims. 
    That is, \psqtool succeeded to perform model checking for a significant 
    time length, even when it timed out, as compared to CBMC that failed to 
    generate the CNF formula, thus did not really test the program for correctness. 
\end{itemize}

For the array search program, the precondition checks that the indexes are in 
range and the postcondition checks whether the return value is consistent. 
For the binary search program both the precondition and the postcondition 
specify the sortedness property for the array. 
For the bubble and selection sort programs, the precondition checks the
sanity of the size of the array, and the postcondition specifies that the 
resulting array holds the same elements of the original array in order. 
For the array partition program, 
the precondition specifies the sanity of the indexes and the size of the array, 
and the postcondition specifies that the resulting array holds the same elements
of the original array partitioned in order left and right of the cursor. 
The linked list insert and remove precondition specifies a non circular, in order, 
list of elements, and the postconditions reassert the same specification with 
an additional cardinality condition. 
The RBBBST insert and remove precondition specifies the full characteristics of 
an RBBBST and that the index arrays involved all hold valid indexes. 
The postconditions reassert the same specification with an additional cardinality
condition. The RBBBST insert also checks the sanity of the newly allocated node
in the tree. 
The memory manager unit has several multidimensional arrays to bookkeep the use 
of memory chucks and the precondition and the postcondition assert the sanity of
the arrays, and the consistency of the module.
We made use of the assume statement \cci{\_\_CPROVER\_assume} and 
loops to implement the preconditions and the quantifiers in CBMC. 

The reduction algorithms used by \psqtool were able to reduce 
the original problems on average to 64.5\% of their memory elements,
and to 51.2\%, and 44.7\% of their logic gate and level counts
in often insignificant time. 
Without these reductions \psqtool timed out on several benchmarks. 
These reductions have no counterparts in CBMC. 

\subsection{SV-COMP 2014 benchmarks}
\label{chap:res:svcompbench}
%% svcompetition benchmarks
Tables~\ref{t:svcomp1} and \ref{t:svcomp2} show the 
results of \psqtool compared against the results reported
in \cite{SVComp2014} for the leading participating 
software verification tools. 
We selected the benchmarks that required no or minor 
syntax modifications for the \psqtool front end 
and compared the execution time of \psqtool with the 
leading tools. 
The tables show the name of the benchmark, its category
and the running time needed by the tools to complete
verification with a timeout of 900 seconds as required
by SVComp 2014 regulations. 

\psqtool succeeded to complete the verification tasks in 
Table~\ref{t:svcomp1} and ranked first or second in terms
of running time. 
\psqtool succeeded to complete the verification tasks 
in Table~\ref{t:svcomp2} where the leading tools either
timed out, reported runtime errors, or reported a wrong 
result (produced a counterexample when the benchmark is 
correct, or produced a claim of correctness when the
benchmark had a known defect). 

For the last five verification tasks in Table~\ref{t:svcomp2}
\psqtool timed out without producing a conclusive result
while some of the leading tools returned the correct 
result, reported a timeout, or reported an inconclusive
result. 

\subsubsection{Interactive sessions}
We inspected the results of the benchmarks
where the automated \psqtool script did not return a conclusive result. 
We were able to verify several of them interactively via trying ABC synthesis 
reduction and verification commands and techniques with different configurations 
and settings. 
%
%We were able to interactively verify several of the 
%benchmarks using \psqtool with manually applying ABC reduction 
%and verification commands with different configurations and 
%settings. 
This is not possible with the tools we compared to.
%except possible
%the CPAChecker tool which allows the application of 
%word level manipulation algorithms. 

%%%%%%%%%%%%%%%%%%%%%%%%%%%%%%%%%%%%%%%%%%%%%%%%%%%%%%
%%%%%%%%%%%%%%%%%%%%%%%%%%%%%%%%%%%%%%%%%%%%%%%%%%%%%%

\section{Related work}
\label{s:related}
%Several tools for program formalization and analysis 
%have emerged in the last few years. 
%In the following section, we will describe what they do
%and compare our method to them. 
%The languages Alloy~\cite{kodkodTJ2007}, Z~\cite{znotation1992}, 
%OCL~\cite{ocl99}, and B~\cite{bbook96} are all 
%declarative languages amenable for validation and checking.

ESBMC~\cite{ESBMC2014TACAS} uses SMT solvers to verify multi-threaded 
C programs by forcing bounds on the number of context switches, loop iterations, and 
recursive calls. It also uses an aligned memory model that resolves pointers. 
We differ in that we use AIG solvers, where we do not need to have bounds for 
loop iterations, and we plan in the future to extend our 
approach to concurrent programs by encoding 
each execution thread as a separate program counter and limiting the number of
program counters. 

VCC~\cite{VCC2009TPHOLS} is an industrial strength verification framework for
for concurrent low level C programs. 
It has a ghost type state system that tracks the validity of memory objects; 
i.e., references to memory objects do not overlap type states.
It generates verification conditions that can be checked by Boogie; i.e. a high order 
logic analyzer. Boogie in turn uses the Z3 SMT solver~\cite{Z309} for automatic
verification and Isabelle~\cite{Isabelle2002} for interactive verification. 
FranklinBIT~\cite{FranklinBit2014} takes a bit-vector program with a verification 
condition and computes an unsound approximation (not over and not under) 
of the verification condition. Then it uses a logic solvers (SMT or propositional)
to decide the original verification condition strengthened with the inductive 
unsound approximation.

LLBMC~\cite{LLBMC-ASE-2013} is a bit precise bounded model checker for low level 
C programs that works on an intermediate assembly representation of the program 
using an SMT solver. 
Ultimate Automizer~\cite{AutomizerCAV2013} computes {\em appropriate} 
abstractions of programs based on statements as alphabet atoms in an 
automata framework. 
The work in \cite{BDD-CPACheker2012} is implemented within the 
CPACheker~\cite{CPACheckerFramework2012} 
platform to verify {\em event condition action } (ECA) systems using BDDs. 
Our method allows the use of symbolic model checking through ABC that makes use
of BDD without the need to restrict ourselves to ECA systems. 
The CPAChecker~\cite{CPACheckerFramework2012} framework passes a program \Pm and
an invariant $\Psi$ to model checkers for verification. 
In case a model checker returns an inconclusive result, CPAChecker uses partial 
results to formulate a state predicate formula $\Phi$ where the invariant holds. 
This incrementally reduces the work of the next model checker to check $\neg \Phi$. 
In the future, we plan to extend the same idea to the ABC model checker. 
BLAST~\cite{BLAST-2007Beye} is the predecessor of CPAChecker and it uses abstraction
and refinement techniques to model check temporal properties of C programs. 
Similar to CPAChecker is UFO~\cite{UFO-AlbarghouthiCAV2012} that 
provides a framework for exploring abstraction and interpolation techniques for 
software verification and uses SMT solvers as a backend. 

CBMC~\cite{cbmcCKL2004} is a bounded model checker for ANSI-C
programs that checks for properties such as pointer safety,
array bounds and user assert statements. Given an ANSI-C program
and a bound on the range of variables, CBMC produces a Boolean 
formula in CNF and checks the formula with SAT solvers. 
CBMC %does not exploit word level transformations and 
relies mostly on the power and speed of SAT solvers. 
%Often times it even fails to generate the CNF formula.
SAT solvers often face an exponential blow up in the number 
of possible assignments to the atomic propositions. 
This problem, known as the {\it state explosion},
and the large number of variables and clauses used in the CNF
encoding, limit the CBMC analysis
to relatively small bounds.% e.g., fewer 
%than 16 entries in an array sort program~\cite{SEBAC07}. 

The work in~\cite{Arm06boundedmodel} uses an encoding similar 
to that of CBMC, but instead of producing a CNF 
formula it produces an SMT formula. 
The SMT formula allows for variables with no range bounds. 
The loops are still unrolled up to a finite bound and
loop completion assertions fire in case the bound imposed on 
the number of loop iterations was small and the 
loop guard was still satisfiable. 
The higher level solver may now decide to 
reproduce a new SMT formula with bigger loop unrolling bounds
and call the SMT solver on the new formula. 
We differ in that our encoding to AIG circuits requires no
bounds on loop iterations. 
This allows for more succinct representation of programs than with 
SMT.

Java PathFinder~\cite{jpfVisser03visser} (JPF),
is a popular explicit-state model checker for Java programs.
It implements a customized Java virtual machine (JVM) that
supports state backtracking and allows programs to check
properties of a wide range of Java programs without the need
to respecify the programs in specialized languages. 
JPF does not apply any program transformations. JPF 
also does not scale well in the presence of loops
and branches with long running time.
JPF operates at the word level, however, it 
does not make use of program specific properties
and exhaustively searches the tree of possible
executions.
We differ in that we operate at the bit level and we apply reduction
techniques that allow backtracking
algorithms to run faster as they operate on a smaller state space. 
%Also isomorphism is easier to detect when the program
%is smaller and thus pruning the search space becomes 
%easier.

%SEM~\cite{sem96} is another tool that performs a search directly
%on the problem without converting it to simpler logic. 
%It uses a basic backtracking search based algorithm 
%coupled with a powerful constraint solver.
%It also uses {\it isomorphism detection} techniques to 
%avoid searching symmetrical executions. 
%SEM is known to perform well on problems with equalities.

%DarwinFM~\cite{darwinfm2007} rewrites sorted first order
%logic clauses and eliminates functional symbols and replace
%them with relations. It decides the resulting formula
%using a first order logic solver rather than a SAT solver.
%DarwinFM solves more problems and scales to bounds an order
%of magnitude higher than MACE and Paradox.
%that example of a , Mace~\cite{mace2008}, 

The work of Xie et al~\cite{Xie03FME} translates software
artifacts into equivalent semantics that are 
model-checkable. It applies compositional rules to the translated
system to build its formal semantics in the context of 
message passing systems. 
We differ in that we use a specific program counter based semantics
with a finitization rule to translate the software artifacts to an AIG, 
a model checkable formalization, where reductions, including compositional 
ones, can be used.

%%%%%%%%%%%%%%%%%%%%%%%%%%%%%%%%%%%%%%%%%%%%%%%%%%%%%%
%%%%%%%%%%%%%%%%%%%%%%%%%%%%%%%%%%%%%%%%%%%%%%%%%%%%%%

\section{Conclusion}
\label{s:conclusion}
We presented \psqtool that takes a program and a pair of specifications, 
translates it into an AIG circuit, reduces the circuit using the 
ABC synthesis reduction techniques, and then check the circuit for correctness
using the ABC verification techniques. 
\psqtool scales to bounds larger than that possible with existing tools, and 
solves verification tasks from the software verification competition benchmarks
that leading tools at the competition did not solve within the specified timeout. 
In the future, we plan to extend \psqtool to verify concurrent programs using 
several program counters to represent multi-threaded executions. 
We also plan to integrate \psqtool with existing frameworks that benefit from word
level optimizations and techniques such as CPAChecker to leverage both word level
and bit level techniques as well as cover an extended syntax set through existing
front ends.

%%%%%%%%%%%%%%%%%%%%%%%%%%%%%%%%%%%%%%%%%%%%%%%%%%%%%%
%%%%%%%%%%%%%%%%%%%%%%%%%%%%%%%%%%%%%%%%%%%%%%%%%%%%%%

\bibliographystyle{ieeetr}
{\footnotesize %{\scriptsize %{\small
\bibliography{ref}
}

\end{document}